\documentclass[review]{elsarticle}
\usepackage{amsmath}
\usepackage{amssymb}
\usepackage{epsfig}
\usepackage{color}
\usepackage{color,soul}
\soulregister\cite7
\soulregister\ref7
\usepackage{textgreek}
\usepackage{float}
\usepackage{caption}    
\usepackage{algorithm}
\usepackage{algorithmic}
\usepackage{ulem}
\usepackage[pdftex]{hyperref}
\usepackage{graphicx}
\usepackage{subcaption}
\usepackage[titletoc,title]{appendix}
\usepackage{geometry}
\definecolor{awesome}{rgb}{1.0, 0.13,0.32}

\begin{document}
\renewcommand\appendix{\par
  \setcounter{section}{0}
  \setcounter{subsection}{0}
  \setcounter{figure}{0}
  \setcounter{table}{0}
  \renewcommand\thesection{Appendix \Alph{section}}
  \renewcommand\thefigure{\Alph{section}\arabic{figure}}
  \renewcommand\thetable{\Alph{section}\arabic{table}}
}

\begin{frontmatter}
\title{A Consistent Spatially Adaptive Smoothed Particle Hydrodynamics Method for Fluid-Structure Interactions}

\author[WISC]{Wei Hu}
\author[WISC]{Guannan Guo}
\author[TU]{Xiaozhe Hu}
\author[WISC]{Dan Negrut}
\author[PNNL]{Zhijie Xu}
\author[WISC]{Wenxiao Pan\corref{cor}}
\ead{wpan9@wisc.edu}

\cortext[cor]{Corresponding author}
\address[WISC]{Department of Mechanical Engineering, University of Wisconsin-Madison, Madison, WI 53706, USA}
\address[TU]{Department of Mathematics, Tufts University, Medford, MA 02155, USA}
\address[PNNL]{Advanced Computing, Mathematics \& Data Division, Pacific Northwest National Laboratory, Richland, WA 99352, USA}

\begin{abstract}
A new consistent, spatially adaptive, smoothed particle hydrodynamics (SPH) method for Fluid-Structure Interactions (FSI) is presented. The method combines several attributes that have not been simultaneously satisfied by other SPH methods. Specifically, it is second-order convergent; it allows for resolutions spatially adapted with moving (translating and rotating) boundaries of arbitrary geometries; and, it accelerates the FSI solution as the adaptive approach leads to fewer degrees of freedom without sacrificing accuracy. The key ingredients in the method are a consistent discretization of differential operators, a \textit{posteriori} error estimator/distance-based criterion of adaptivity, and a particle-shifting technique. The method is applied in simulating six different flows or FSI problems. The new method's convergence, accuracy, and efficiency attributes are assessed by comparing the results it produces with analytical, finite element, and consistent SPH uniform high-resolution solutions as well as experimental data.  
\end{abstract}


\end{frontmatter}

\section{Introduction}\label{sec:Introduction}

The FSI problem is encountered in a broad range of science and engineering applications such as complex fluids, additive manufacturing, and carbon dioxide capture. FSI problems are as ubiquitous as challenging. Their solution must account for moving fluid-solid interfaces of complex geometries and hydrodynamic interactions between many immersed solid bodies. In this work, we propose a meshless particle method for solving FSI problems, which is based on a spatially adaptive, consistent SPH method. Different from the standard SPH method \cite{Monaghan_SPH_2005}, in which the spatial discretization lacks consistency and hence results in less than second-order convergence that only can be recovered by taking increasingly large number of neighbors \cite{Fatehi2011error}, the consistent SPH method ensures second-order consistency in its discretization of differential operators.

SPH solves partial differential equations (PDEs) numerically by approximating a variable at collocation points (or particles) throughout the domain via a smoothing function with a compact support (or smoothing length). As there is no assumed structure in the spatial distribution of the particles, they are completely \textit{unstructured}. This unstructured nature allows for moving the particles with the dynamics of the problem and hence solving the underlying physics in a Lagrangian sense. As a result, there is no need to discretize the nonlinear advection term when solving the Navier-Stokes equation for fluid dynamics. Due to its meshless, Lagrangian nature, SPH has been widely applied in modeling multiphase flows with moving interfaces \cite{Adami2010surface,Tofighi2013numerical,Tart2016JCP}, complex fluids  \cite{PanJG2012,PanJCP2013,SPHPolymer2016,SPHNoncolloid2016}, materials subject to large deformation \cite{Gray2001elastic,Pan2013material,Hu2016dynamic}, FSI problems  \cite{Hu2014dynamic,Schorgenhumer2013interaction,Pazouki2014FSI}, and various transport phenomena \cite{Pan_LLNS_2014,Tart2015CompGeo,PanBMC2015,SPHProton2015}. 

In most SPH simulations, a uniform resolution is employed throughout the computational domain with SPH particles of uniform spacing and mass. This approach can be prohibitively expensive in modeling FSI problems when the near-field variables (velocity/pressure) surrounding solid bodies or within narrow gaps between solid boundaries must be accurately resolved. In such cases, the expectation is that resolving only a limited number of subdomains at high resolution while maintaining the rest of the simulation domain at lower particle resolution will yield efficiency gains without compromising the accuracy. This idea has been pursued in several multi-resolution or adaptive-resolution SPH methods~\cite{Lastiwka2009,Oger2006wedge,Feldman2007refinement,Lopez2013refinement,Vacondio2013variable,Barcarolo2014refinement,Bian2015DDSPH}. 

In the multi-resolution scenario, the computational domain is pre-partitioned into a couple of subdomains. Each subdomain is discretized with SPH particles of a predefined spacing. What differentiates the methods in this scenario is how the subdomains of different resolutions are coupled, and how the predefined configuration of particle spacing is maintained as SPH particles advect with flow during the simulation. For example, in \cite{Vacondio2013variable,Barcarolo2014refinement,Vacondio2016variable}, each subdomain is assigned its own smoothing length according to its particle spacing. With the inconsistent SPH discretization, the accuracy and continuity of the numerical solution at and near the interface of subdomains is hard to maintain. The errors became notable as the ratio of the two resolutions increases \cite{hu2017consistent}. To tackle this issue, a large, uniform smoothing length was used for all subdomains of different resolutions in \cite{Omidvar2012variable2D}. This enhanced the accuracy attribute of the solution to some degree but led to a large number of neighbors for the particles in the high-resolution subdomain. Alternatively, an approach promoting different-resolution subdomains connected via an overlap region was proposed in \cite{Bian2015DDSPH}. The solutions in the two subdomains were synchronized at each time step via the overlap region where the state information and fluxes were exchanged by an iterative process. For the sake of numerical accuracy, a minimum thickness of the overlap region must be ensured and determined numerically. Two concerns regarding this approach are: 1) bringing the solutions of two subdomains in agreement via an iterative algorithm is costly and ad-hoc; and, 2) if the two subdomains are moving dynamically, continually establishing an adaptive overlap region is challenging. Against this backdrop, we proposed a different multi-resolution SPH method in \cite{hu2017consistent}. Therein, each subdomain still has its own smoothing length, but the consistent discretization of differential operators was enforced via local correction matrices computed at each particle \cite{Fatehi2011error,Traska2015second-order,Pan2017Modeling}. Without resorting to an overlap region, the continuity and second-order accuracy of the numerical solution were guaranteed. Owing to its higher-order accuracy, this multi-resolution SPH method required fewer number of particles to achieve the same accuracy and thereby led to further computational savings. As SPH particles advect with flow and move from one subdomain into the other, the predefined multi-resolution configuration was maintained via dynamically splitting or merging SPH particles. More specifically, a ``big'' particle was split into several ``small'' ones according to a certain geometric arrangement centered at the location of the big particle. Conversely, a group of small particles were merged into a big one at the center-of-mass of these small particles. This  particle splitting/merging technique is similar to those used in literature \cite{Meglicki19933d,Kitsionas2002splitting,Kitsionas2007clump,Feldman2007refinement,Vacondio2012split,Lopez2013refinement,Vacondio2013variable,Barcarolo2014refinement,Vacondio2016variable}. 

The limitation of these multi-resolution methods is that spatial resolutions are not varied adaptively throughout the domain but are predefined for each subdomain. As shown in our previous work, the convergence of numerical solutions is affected by the ratio of resolutions in two adjacent subdomains. To maintain the second-order convergence, this ratio must be less than 2. As a result, the computational savings associated with using multi-resolution methods are limited. Hence, a method with a continuously adaptive spatial resolution is more attractive. Efforts along this line have been attempted in the past. For instance, Oger et al. used adaptive smoothing lengths in the solution domain according to the distances of fluid particles to the solid boundary, with larger smoothing lengths used for particles further away from the solid boundary \cite{Oger2006wedge}. By using spatially adaptive smoothing lengths, they were able to capture the transient dynamics of a wedge moving in water with reduced computational cost \cite{Oger2006wedge}. However, in this approach the smoothing lengths were only allowed to vary within a rather small range (less than 3\% for any two adjacent layers) and did not change during the simulation after initially assigned. Alternatively, in \cite{Shapiro1996adaptive,Owen1998adaptive,Liu2005meshfree,Liu2006adaptive}, SPH particles with ellipsoidal smoothing kernels were used to discretize the computational domain. In two dimensional problems, the two major smoothing lengths of the ellipsoidal kernel were allowed to adaptively vary on different particles according to the local number density of SPH particles. Note that both the method in \cite{Oger2006wedge} and the ellipsoidal-kernel approach were based on the inconsistent SPH discretization, and their accuracy and convergence were not systematically assessed. 

In this work, we propose a new, consistent adaptive-resolution SPH method for modeling FSI. In this method, spatial resolutions in the fluid domain are adaptively varied according to appropriate criteria of adaptivity. In particular, two criteria of adaptivity are assessed. The first one is distance-based -- the spatial resolutions gradually vary according to the distances to the solid boundary. The second one draws on a \textit{posteriori} error estimation. Note that in the literature of adaptive finite element methods (AFEM) for solving second-order elliptic PDEs, there are mainly two types of a \textit{posteriori} error estimators. One is the residual-based error estimator~\cite{Verfurth.R1996a,Nochetto.R;Siebert.K;Veeser.A2009}; the other one is the recovery-based error estimator~\cite{Zienkiewicz.O;Zhu.J1992c,RecoverError_Book2000}.  Although a rigorous convergence theory for AFEM can be established based on the residual-based error estimator~\cite{Nochetto.R;Siebert.K;Veeser.A2009}, we adopt the recovery-based error estimator for the SPH method for the following two reasons. First, the residual-based error estimator is derived from the variational formulation and, therefore, strongly depends on the problem being solved. The recovery-based error estimator is solely based on the computed solution and does not depend on the specific governing PDEs, which makes it more flexible and suitable for the SPH method compared with residual type error estimators. Second, although the theoretical justification of a recovery-based \textit{posteriori} error estimator is still under investigation~\cite{Xu.J;Zhang.Z2004,RecoverError_Zhang2005}, in practice, it is usually more reliable and efficient compared with the residual-based error estimators; i.e., it provides more accurate estimations of the true errors~\cite{RecoverError_Zhang2004,Xu.J;Zhang.Z2004}. 

Against this backdrop, our recovery-based a \textit{posteriori} error estimator for the SPH method borrows from its AFEM counterpart. It measures the difference between the direct and recovered velocity gradient \cite{RecoverError_Book2000}. With this error estimator evaluated at all SPH particles, following the error equidistribution strategy, the particles with larger or smaller errors are marked for refinement or coarsening. A particle shifting technique is then employed to refine or coarsen particle spacings. Thus, a particle is shifted according to the shifting vector computed, and after shifting, hydrodynamic variables are corrected according to its new positions using a second-order interpolation. The two-way coupling of FSI is realized by imposing the no-slip boundary condition (BC) for the fluid at the solid boundary and moving the solid body with the force and torque exerted by the fluid. The forces and torques contributed by the pressure and viscous stress of the fluid are evaluated on the solid boundaries with the consistent discretization. The no-slip BC for the fluid velocity is imposed through ghost solid particles with velocities linearly extrapolated from the fluid velocity \cite{TakedaSPHBC1994,Morris1997modeling,Holmes2011Smooth}, which ensures the second-order accuracy at the boundaries \cite{Macia2011SPHBCs}. The proposed FSI method is validated and demonstrated to reduce computational costs without compromising accuracy. Compared with its mesh-based counterparts; i.e., AFEM and adaptive finite volume methods, the proposed SPH meshless solution for \textit{moving} boundary FSI problems allows for adaptively varying spatial resolutions without the need of remeshing, a task that is generally expensive and unscalable. 

This paper is organized as follows. Section \ref{sec:method} describes the proposed consistent adaptive-resolution SPH method. Starting with a consistent SPH discretization, we explain next the approach for dynamically adapting spatial resolutions in solving the time-dependent problems. In Section \ref{sec:Simu_results}, we demonstrate the accuracy, convergence, and efficiency of the proposed method through numerical tests. We model six different flows -- the transient Poiseuille flow, flow around a periodic array of cylinders, flow around a cylinder in a narrow channel, two colloids rotating under a shear flow, dynamics of a low-symmetry active colloid, and collective dynamics of a suspension of active colloids. The numerical results obtained are compared with analytical predictions, experimental data, and simulation results from the consistent uniform-resolution SPH or from other numerical methods such as FEM. Finally, we conclude and suggest directions of future work in Section \ref{sec:conclusion}.


\section{Consistent adaptive-resolution SPH method} \label{sec:method}

\subsection{Governing equations} \label{section:GovEqu}

In FSI problems, the computational domain can be represented as $\Omega=\Omega_f\cup\Omega_s$, where $\Omega_f$ and $\Omega_s$ are the sub-domains occupied by the fluid and solid, respectively. The boundary $\Gamma$ separates $\Omega_f$ and $\Omega_s$, i.e., $\Gamma = \Omega_f \cap \Omega_s$. 

\subsubsection{Lagrangian hydrodynamics} \label{subsection:NS}

The fluid velocity $\mathbf{v}$ and pressure $p$ are governed by the continuity and Naiver-Stokes (NS) equations: 
\begin{equation}
\begin{cases}
\frac{d\rho}{d t}=-\rho \: \nabla \cdot \textbf{v} \\
\frac{d \mathbf{v}}{d t} = -\frac{\nabla p}{\rho} + \nu \nabla^2 \mathbf{v} + \mathbf{g} ~~~~~~\text{for}~~~ \textbf{x}\in \Omega_f \; ,
\label{equ:NS}
\end{cases}
\end{equation}
where $\mathbf{g}$ denotes the body force per unit mass such as gravity; $\nu = \frac{\mu}{\rho}$ is the kinematic viscosity of the fluid; $\rho$ and $\mu$ are the fluid density and viscosity, respectively. Here, the NS equations are written from a Lagrangian point of view with $d/dt=\partial/\partial t + \mathbf{v} \cdot \nabla$ denoting the total time derivative. Eq. \eqref{equ:NS} is closed by the equation of state (EOS): 
\begin{equation}\label{equ:state-equ}
p= c^2 \rho - p_0 \; ,
\end{equation}
with a low March number ($<0.1$) to approximately model an incompressible fluid. Here, $c$ is the speed of sound and $p_0$ is the background pressure, which can guarantee a non-negative pressure in the flow field to make the simulation stable \cite{Balsara1995StaSPH,Swegle1995StaSPH,Morris1996StaSPH}.

\subsubsection{Fluid-structure interactions} 
\label{subsec:FSI}

The fluid-solid coupling is modeled by imposing a no-slip BC for the fluid at the solid boundary; and, for the solid body, by accounting for the force and torque exerted by the fluid. The no-slip BC assumes the generic form $\textbf{v}({\bf x})=\textbf{v}_B({\bf x})$; i.e., at any point of the boundary ${\bf x \in \Gamma}$, the velocity ${\bf v}$ of the fluid and ${\bf v}_B$ of the solid are identical. This condition should hold regardless whether the solid is rigid or deformable. For a rigid body, this condition can be further specified to read:
\begin{equation}\label{equ:noslipBC}
\textbf{v} = \textbf{v}_B = \mathbf{v}_d + \mathbf{r} \times \boldsymbol{\omega}_d ~~~~~~\text{for}~~~ \textbf{x}\in \Gamma \; ,
\end{equation}
where $\mathbf{v}_d$ and $\boldsymbol{\omega}_d$ denote the linear and angular velocities of the solid (rigid) body, respectively, and $\mathbf{r}$ denotes the vector from the center-of-mass of the rigid body to the location ${\bf x} \in \Gamma$. Although the methodology presented here applies equally well to rigid and deformable bodies, the presentation continues with an assumption that the solid body is rigid (infinitely stiff). Its equations of motion are then governed by the differential equations:
\begin{equation}
\begin{cases}
m_d \frac{d\mathbf{v}_d}{d t}= \mathbf{F}_d \\
I_d \frac{d\boldsymbol{\omega}_d}{d t} = \mathbf{T}_d \; ,
\label{equ:rigidbody}
\end{cases}
\end{equation}
where $m_d$ and $I_d$ represent the total mass and moment of inertia of the solid body, respectively. The total force $\mathbf{F}_d$ and torque $\mathbf{T}_d$ exerted on the solid body by fluid can be computed as:
\begin{equation}\label{equ:force_torque}
\begin{cases}
\mathbf{F}_d = \int_{\Gamma} \mathbf{n}\cdot \boldsymbol{\sigma} d\textbf{x} \\
\mathbf{T}_d = \int_{\Gamma} \mathbf{r} \times (\mathbf{n} \cdot  \boldsymbol{\sigma})d\textbf{x} \\
\boldsymbol{\sigma} = -p \mathbf{I} + \boldsymbol{\tau}~~\text{with}~~ \boldsymbol{\tau} = \mu \left [\nabla \mathbf{v} +(\nabla \mathbf{v})^{\mathrm{T}}\right ] \; ,
\end{cases}
\end{equation}
where $\boldsymbol{\sigma}$ is the stress tensor; $\boldsymbol{\tau}$ is the viscous stress.

\subsection{Spatial discretization}
\label{subsec:SPH Cons_Dis}  

In SPH, the computational domain is discretized by a set of particles. The particles within $\Omega_f$ and $\Omega_s$ are referred to as \lq\lq{}fluid\rq\rq{} and \lq\lq{}solid\rq\rq{} particles, respectively. The value of a function $f$ at a particle $i$ is then approximated as \cite{Monaghan_SPH_2005}:
\begin{equation}\label{equ:sph_fun}
f_i = \sum\limits_{j}f_j W_{ij}\mathcal{V}_j \; .
\end{equation}
Here, $\mathcal{V}$ is the volume of a SPH particle defined as $\mathcal{V}_i = (\sum\limits_{j}W_{ij})^{-1}$. Hence, the mass of a SPH particle is given by $m_i = \rho_i \mathcal{V}_i$.
In Eq. \eqref{equ:sph_fun}, $W_{ij} = W(\textbf{r}_{ij})$ is the kernel function. In this work, we use the quintic spline function $W$ that has continuous and smooth first and second derivatives:
\begin{equation}\label{equ:sph_ker}
W_{ij} = \alpha_d
\begin{cases}
\ (3-R)^5 - 6(2-R)^5 + 15(1-R)^5 ~~~0 \leqq R < 1 \\
\ (3-R)^5 - 6(2-R)^5 ~~~~~~~~~~~~~~~~~~~~~1 \leqq R < 2 \\
\ (3-R)^5 ~~~~~~~~~~~~~~~~~~~~~~~~~~~~~~~~~~~~~2 \leqq R < 3 \\
\ \ 0 ~~~~~~~~~~~~~~~~~~~~~~~~~~~~~~~~~~~~~~~~~~~~~~~~~~~ R \geqq 3. \\
\end{cases} \\
\end{equation}
In Eq. \eqref{equ:sph_ker}, the constant $\alpha_d$ assumes the value of $7/478\pi h^2$ for two-dimensional (2D) simulations; $R=\frac{r_{ij}}{h}$ with $h$ denoting the kernel length and $r_{ij}=|\textbf{r}_{ij}|$. Here, $\textbf{r}_{ij} = \textbf{x}_i - \textbf{x}_j$ where $\textbf{x}_i$ is the position of particle $i$. For future reference, $\textbf{e}_{ij}=\textbf{r}_{ij}/r_{ij}$. With this kernel function, an SPH particle $j$  contributes to the summation in Eq. \eqref{equ:sph_fun} for approximating the function at particle $i$ only when $j \in {\cal{N}}_{h,i} = \left\{ \textbf{x}_j~ :~ |\textbf{x}_j - \textbf{x}_i| < 3h \right\}$. 
 
In the consistent SPH discretization \cite{hu2017consistent,Fatehi2011error,Traska2015second-order,Pan2017Modeling}, the gradient and Laplacian of the function $f_i$ are discretized as:
\begin{equation}\label{equ:gra_ope}
\nabla f_{i}=\sum\limits_j (f_{j}-f_{i})\left(\textbf{G}_{i}\cdot \nabla_i W_{ij}\right)\mathcal{V}_{j},
\end{equation}
\begin{equation}\label{equ:lap_ope}
\nabla^{2} f_{i}=2\sum\limits_j [ \textbf{L}_{i}:(\textbf{e}_{ij} \otimes \nabla_i W_{ij})] (\frac{f_{i}-f_{j}}{r_{ij}} - \textbf{e}_{ij}\cdot\nabla f_{i}) \mathcal{V}_{j},
\end{equation}
where ``$\otimes$" represents the dyadic product of two vectors, and ``$:$" represents the double dot product of two matrices. Previous studies have demonstrated that this discretization of the gradient and Laplacian, which involves the symmetric correction matrices $\textbf{G}_i$, $\textbf{L}_i \in {\mathbb{R}}^{2\times 2}$, guarantees the exact gradient for linear functions and Laplacian for parabolic functions regardless of the ratio of $h/\Delta x$ and hence displays second-order accuracy \cite{Fatehi2011error}. Here, $\Delta x$ is the particle spacing. A detailed description of $\textbf{G}_i$ and $\textbf{L}_i$ can be found in \cite{hu2017consistent}. For a vector function, the discretization in Eq. \eqref{equ:gra_ope} is used for defining the discrete divergence operator.  Hence, the consistent SPH discretization of the governing equations is given as:
\begin{equation}\label{equ:continuity_discretization}
\frac{d\rho_i}{dt}= -\rho_i \: \sum\limits_j (\textbf{v} _{j}- \textbf{v}_{i}) \cdot \left(\textbf{G}_{i} \cdot \nabla_i W_{ij}\right)\mathcal{V}_{j}\; ,
\end{equation}
\begin{equation}\label{equ:NS_discretization}
\frac{d\mathbf{v}_i}{dt} = -\sum\limits_j \frac{1}{\rho_j} (p_{j}-p_{i})\left(\textbf{G}_{i}\cdot \nabla_i W_{ij}\right )\mathcal{V}_{j} + 2\sum\limits_j \nu_j [ \textbf{L}_{i}:(\textbf{e}_{ij} \otimes \nabla_i W_{ij})]\\
                 (\frac{\mathbf{v}_{i}-\mathbf{v}_{j}}{r_{ij}} - \textbf{e}_{ij}\cdot\nabla \mathbf{v}_{i}) \mathcal{V}_{j} + \mathbf{g}_i \; .
\end{equation}

As noted, the consistent SPH discretization does not guarantee local conservation of momentum. As a result, the total force exerted by the fluid on the solid can not be computed directly from the summation of forces contributed by the solid particles to fluid particles as in the conservative SPH methods \cite{bian2014splitting}. Alternatively, we compute the force from the pressure and viscous stress evaluated on the solid boundary as in Eq. \eqref{equ:force_torque}. To that end, we assign a new set of  particles exactly on the boundary surface. Note that these ``surface" particles do not participate in discretizing and solving the governing equations but only serve for evaluating the pressure and viscous stress exerted by the fluid on the solid boundary, from which the total force and torque on the solid are obtained. For evaluating the viscous stress, the gradient operator in $\boldsymbol{\tau}=\mu \left [\nabla \mathbf{v} +(\nabla \mathbf{v})^{\mathrm{T}}\right ]$ is discretized on the surface particles according to Eq. \eqref{equ:gra_ope} with only fluid particles included in the summation as the neighbor particles. The pressure on the surface particles is calculated as \cite{Adami2012generalized}: 
\begin{equation}
p_i =  \frac{\sum\limits_{j\in \text{fluid}} p_j W_{ij} + (\textbf{g} - \textbf{f}_i)\cdot \sum\limits_{j\in \text{fluid}}\rho_j \textbf{r}_{ij}W_{ij}}
            {\sum\limits_{j\in \text{fluid}} W_{ij}} \; ,
\end{equation}
which considers the contribution of the net force acting on the solid from the body force ($\textbf{g}$) of the fluid and acceleration ($\textbf{f}_i$) of the surface particles. The latter is defined as:
\begin{equation}
\textbf{f}_{i} = \frac{d\mathbf{v}_d}{d t} + \frac{d\boldsymbol{\omega}_d}{d t}\times \mathbf{r}_i + \boldsymbol{\omega}_d \times (\boldsymbol{\omega}_d \times \mathbf{r}_i)
\end{equation}
Finally, the total force and torque applied on the immersed solid body can be evaluated by approximating Eq. \eqref{equ:force_torque} using the midpoint quadrature rule:
\begin{equation}
\textbf{F}_{d} = \sum\limits_{i\in \text{surface}} (\textbf{n}_i \cdot \boldsymbol{\sigma}_i) \Delta S ~~~\text{and}~~~ \mathbf{T}_d = \sum\limits_{i\in \text{surface}} \mathbf{r}_i \times (\mathbf{n}_i \cdot  \boldsymbol{\sigma}_i)\Delta S \; ,
\end{equation}
where $\textbf{n}_i$ is the exterior normal vector at the position of surface particle $i$; $\Delta S$ is the spacing between two neighbor surface particles; and $\mathbf{r}_i$ denotes the vector from the center-of-mass of the solid body pointing to the surface particle $i$.

\subsection{Imposition of the no-slip boundary condition}\label{subsec:Bou_Cons}  
To accurately impose the no-slip boundary condition for the fluid velocity, the SPH approximations of velocity and its spatial derivatives for the fluid particles near the fluid-solid boundary must attain full support of the kernel contained in the domain ($\Omega_f \cup \Omega_s$). Thus, we follow the approach used in our previous work \cite{Pan2017Modeling,hu2017consistent} and in literature \cite{TakedaSPHBC1994,Morris1997modeling,Holmes2011Smooth} to lay several layers of ghost particles in the solid domain near the boundary. These ghost particles are assigned velocities linearly extrapolated from the velocities of fluid particles, i. e.,
\begin{equation} \label{equ:linear_extrapol}
\textbf{v}_j = \frac{d_{j}}{d_{i}}(\textbf{v}_B - \textbf{v}_i) + \textbf{v}_{B} \; ,
\end{equation}
where $i$ is a fluid particle; $j$ represents a solid ghost particle; $d_{i}$ and $d_{j}$ denote the closest perpendicular distances to the boundary for the fluid and ghost particles, respectively; and $\textbf{v}_{B}$ is the velocity of the solid boundary as in Eq. \eqref{equ:noslipBC}. This linear extrapolation introduces a local $\mathcal{O} (h^2)$ error \cite{Macia2011SPHBCs} matching the second-order accuracy of the consistent SPH. In practice, $d_i$ and $d_j$ in Eq. \eqref{equ:linear_extrapol} are approximated by \cite{Holmes2011Smooth}: 
\begin{equation*}
d_i = \omega h_{i} (2\chi_i-1) \; , \qquad \qquad
d_j = \omega h_{j} (2\chi_j-1) \; ,
\end{equation*}
with $\omega=3$ for the quintic spline function used herein. The indicator $\chi$ used to differentiate fluid and solid particles is given as:
\begin{equation*}
\chi_i = \frac{\sum\limits_{k\in \mathrm{fluid} }W_{ik}}{\sum\limits_{k}W_{ik}} \; , \qquad \qquad
\chi_j = \frac{\sum\limits_{k\in \mathrm{solid} }W_{jk}}{\sum\limits_{k}W_{jk}}.
\end{equation*}

\subsection{Time integration}\label{subsec:Time_Integ}  

The variables defined on fluid particles are updated via a standard second-order, explicit predictor-corrector scheme  \cite{Monaghan1989problem,Monaghan1994simulating,hu2017consistent}. In the predictor step, the intermediate velocity $\bar{\textbf{v}}_{i}$ and position $\bar{\textbf{x}}_{i}$ at the intermediate time step ($t+\frac{\Delta t}{2}$) are first predicted as:
\begin{equation*}
\begin{cases}
\ \bar{\textbf{v}}_{i}({t+\frac{\Delta t}{2}})= \textbf{v}_{i}(t) + \frac{\Delta t}{2} \textbf{a}_{i}(t)  \\
\ \bar{\textbf{x}}_{i}({t+\frac{\Delta t}{2}})= \textbf{x}_{i}(t) + \frac{\Delta t}{2} \textbf{v}_{i}(t) \; , \\
\end{cases} \\
\end{equation*}
where $\textbf{a}_{i} = \frac{d\mathbf{v}_i}{dt}$ is the total acceleration of particle $i$ determined from Eq. \eqref{equ:NS_discretization}. The intermediate density $\bar{\rho}_{i}(t+\frac{\Delta t}{2})$ and pressure $\bar{p}_{i}(t+\frac{\Delta t}{2})$ are then obtained via Eqs.~\eqref{equ:continuity_discretization} and \eqref{equ:state-equ}, respectively. After the predictor step, $\textbf{a}_{i}(t+\frac{\Delta t}{2})$ is evaluated and subsequently used in the corrector step:
\begin{equation*}
\begin{cases}
\ \textbf{v}_{i}(t+\frac{\Delta t}{2})= \textbf{v}_{i}(t) + \frac{\Delta t}{2} \textbf{a}_{i}(t+\frac{\Delta t}{2}) \\
\ \textbf{x}_{i}(t+\frac{\Delta t}{2})= \textbf{x}_{i}(t) + \frac{\Delta t}{2} \textbf{v}_{i}(t+\frac{\Delta t}{2}). \\
\end{cases} \\
\end{equation*}
Finally, the particle's velocity and position at $t+\Delta t$ are updated by: 
\begin{equation*}
\begin{cases}
\ \textbf{v}_{i}(t+\Delta t) = 2\textbf{v}_{i}(t+\frac{\Delta t}{2}) - \textbf{v}_{i}(t)  \\
\ \textbf{x}_{i}(t+\Delta t) = 2\textbf{x}_{i}(t+\frac{\Delta t}{2}) - \textbf{x}_{i}(t) \; , \\
\end{cases} \\
\end{equation*}
while the density $\rho_{i}(t+\Delta t)$ and pressure $p_{i}(t+\Delta t)$ are obtained via Eqs.~\eqref{equ:continuity_discretization} and \eqref{equ:state-equ}, respectively. In this predictor-corrector scheme, the time step $\Delta t$ is constrained by the CFL condition \cite{Monaghan1992smoothed}, magnitude of acceleration $\left | \textbf{a}_i \right |$, and viscous dispersion as: 
\begin{equation*}
\Delta t \leq \min \{ 
0.25 \frac{h}{c},~ 
0.25 \min_i(\frac{h}{\left | \textbf{a}_i \right |})^{\frac{1}{2}},~
0.125 \frac{h^{2}}{\nu}
\} \;.
\end{equation*}

For the solid body, the velocity $\textbf{v}_{d}$, angular velocity $\boldsymbol{\omega}_d$, angle $\boldsymbol{\theta}_d$ and position $\textbf{x}_{d}$ at the intermediate time step are first computed from:
\begin{equation*}
\begin{cases}
\ \textbf{v}_{d}(t+\frac{\Delta t}{2})= \textbf{v}_{d}(t) + \frac{\Delta t}{2} \frac{\textbf{F}_{d}(t)}{m_d} \\
\ \textbf{x}_{d}(t+\frac{\Delta t}{2})= \textbf{x}_{d}(t) + \frac{\Delta t}{2} \textbf{v}_{d}(t) \\
\ \boldsymbol{\omega}_{d}(t+\frac{\Delta t}{2})= \boldsymbol{\omega}_{d}(t) + \frac{\Delta t}{2} \frac{\textbf{T}_{d}(t)}{I_d} \\
\ \boldsymbol{\theta}_{d}(t+\frac{\Delta t}{2})= \boldsymbol{\theta}_{d}(t) + \frac{\Delta t}{2} \boldsymbol{\omega}_{d}(t). \\
\end{cases} \\
\end{equation*}
With this information at the intermediate time step, $\textbf{F}_{d}(t+\frac{\Delta t}{2})$ and $\textbf{T}_{d}(t+\frac{\Delta t}{2})$ are evaluated and used to update the variables of the next time step as:
\begin{equation*}
\begin{cases}
\ \textbf{v}_{d}(t+\Delta t)= \textbf{v}_{d}(t) + \Delta t \frac{\textbf{F}_{d}(t+\frac{\Delta t}{2})}{m_d} \\
\ \textbf{x}_{d}(t+\Delta t)= \textbf{x}_{d}(t) + \Delta t \textbf{v}_{d}(t+\frac{\Delta t}{2}) \\
\ \boldsymbol{\omega}_{d}(t+\Delta t)= \boldsymbol{\omega}_{d}(t) + \Delta t \frac{\textbf{T}_{d}(t+\frac{\Delta t}{2})}{I_d} \\
\ \boldsymbol{\theta}_{d}(t+\Delta t)= \boldsymbol{\theta}_{d}(t) + \Delta t \boldsymbol{\omega}_{d}(t+\frac{\Delta t}{2}). \\
\end{cases} \\
\end{equation*}

\subsection{Approach for adapting spatial resolutions}\label{sec:adaptive}

In the present work, we assessed two spatial-resolution adaptivity criteria. The first one is distance-based, and it varies the spatial resolution in the fluid domain gradually according to distances to the solid boundary. The second one is based on an error estimator, namely, the recovery-based a \textit{posteriori} error estimator.

\subsubsection{Recovery-based error estimator}\label{sec:Err_Cri}

The recovery-based error estimator measures the difference between the direct and recovered velocity gradient in the energy norm \cite{RecoverError_Book2000} and has been commonly used in the adaptive FEM \cite{RecoverError_Zhang2004,RecoverError_Zhang2005}. In the framework of consistent SPH, this error estimator can be evaluated locally on each fluid particle $i$ at every timestep as:
\begin{equation}\label{equ:recover_error}
\| \mathbf{e}_i \|^2 = m_i \| (\mathbf{R}[\mathbf{v}]_i -\nabla \mathbf{v}_i) \| ^2 \; .
\end{equation}
Here, $\mathbf{R}[\mathbf{v}]_i$ represents the recovered velocity gradient
\begin{equation}\label{equ:recover_gradient}
\mathbf{R}[\mathbf{v}]_i = \sum\limits_{j} (\nabla \mathbf{v}_j) W_{ij} \mathcal{V}_j \; ,
\end{equation}
where $\nabla \mathbf{v}_i$ is the local velocity gradient evaluated in the discrete setting using Eq. \eqref{equ:gra_ope}. The energy norm $\| ~ \|^2$ in Eq. \eqref{equ:recover_error} is defined as the element-wise inner product, i.e., $\| \mathbf{e} \|^2 = \mathbf{e}: \mathbf{e} = e^{mn} e^{mn}$, where the Einstein summation convention is adopted.

\subsubsection{Adaptive algorithm}\label{sec:Adapt_Algo}  
Initially, the SPH particles are distributed with a uniform resolution (spacing) $\Delta x_{ini}$ and mass $m_0$. During the simulation, the initial spacing is dynamically adapted to new distributions of variable resolution. In this process, the total number of SPH particles ($N_{tot}$) and, therefore, the total degree of freedom (DOF) count, do not change. Before the procedure starts, we preset the finest resolution $\Delta x_H = \Delta x_{ini}/n$ allowed in the simulation domain. Given that, the mass of the smallest particle is $m_H=m_0/n^2$ in 2D simulations. With the total number and mass of SPH particles conserved, the mass of the largest particle is set as $m_L=(2n^2-1)m_0/n^2$. The maximum resolution ratio $\Upsilon_{max}$ is defined as the ratio of the coarsest resolution ($\Delta x_L$) to the finest resolution ($\Delta x_H$) throughout the simulation domain, i.e., $\Upsilon_{max} = {\Delta x_L}/{\Delta x_H}$. With the masses of largest and smallest particles determined, we obtain $\Upsilon_{max} = \sqrt{2n^2-1}$. Thus, if a smaller $\Delta x_H$ is chosen, $\Delta x_L$ and $\Upsilon_{max}$ are larger. Given the simulation domain is finite, a larger $\Upsilon_{max}$ results in larger ratios of resolutions of any two adjacent SPH particles. Here, we choose $\Delta x_H$ or $n$ such that the maximum ratio of resolutions of any two adjacent SPH particles is less than 2. Otherwise, according to our previous study \cite{hu2017consistent}, the convergence of the solution could worsen. 

At the end of each time step, after all variables have been updated following the schemes described in Sections \ref{subsec:SPH Cons_Dis}--\ref{subsec:Time_Integ} based on the current spatial resolutions, a three-step process is carried out to adjust the spatial resolutions as follows:
$$
\textbf{ESTIMATE}  \rightarrow \textbf{MARK}  \rightarrow \textbf{REFINE/COARSEN}.
$$

In the \textbf{ESTIMATE} step, depending on which criterion is employed; i.e., error-based or distance-based, we first calculate for each fluid particle either its recovered error of the velocity gradient, or its distance to specified solid boundaries. 

\begin{enumerate}
\item \noindent  If the \textit{error-based} criterion is employed:

\noindent \textbf{The MARK step}: The local error estimated via Eq. \eqref{equ:recover_error} in the ESTIMATE step will indicate which particles are relatively too large or too small in the current particle spatial distribution. These particles are then MARK-ed for refining or coarsening. Following the error equidistribution strategy, particles with larger or smaller errors are potentially candidates for refinement or coarsening. More specifically, the top $5\%$ of particles with the largest errors will be refined if their current resolutions have not reached the preset $\Delta x_H$; and, the bottom $5\%$ of particles with the smallest errors will be coarsened if their resolutions have not reached the predetermined $\Delta x_L$. The 5\%-value is conservative enough to ensure the numerical stability.

\noindent \textbf{The REFINE/COARSEN step}: The refinement for the error-based criterion is realized by reducing the mass of a SPH particle $m_i$ by $\Delta m$; and, the coarsening is enforced by increasing $m_i$ by $\Delta m$. In our study, we found that the simulation is more stable when $\Delta m = 0.1\%m_0$. For the particle whose mass is changed, its smoothing length $h_i$ must be changed proportionally to the new mass assigned, i.e., by the factor of $\sqrt{\frac{m_i\pm \Delta m}{m_i}}$ (in 2D simulations). 

\item If the \textit{distance-based} criterion is employed:

\noindent \textbf{The MARK step}: The fluid particles are sorted according to their distances to specified solid boundaries calculated in the ESTIMATE step and all MARK-ed for refining or coarsening. As the distance of each fluid particle to the solid boundary does not change much between two continuous time steps, the refining/coarsening needed is reasonably small. 

\noindent \textbf{The REFINE/COARSEN step}: According to the sorting of distances to the solid boundary, all fluid particles are assigned new masses and kernel lengths. In a technique that renders the mass of particles closer to the solid boundary smaller, the refining starts with a step in which we preset the number of smallest particles ($N_H$) needed to resolve the subdomain near the solid boundary. With the total number and mass of SPH particles conserved, we next solve for an integer $N_m$ that best approximates the nonlinear equation:
\begin{equation}\label{equ:preadapt_Nm}
N_H m_H + (N_{tot}-N_H-N_m) m_L + \sum_{i=1}^{N_m} m_H \left (\frac{m_L}{m_H} \right )^{\frac{i}{N_m+1}}  = N_{tot} m_0 \; .
\end{equation}
With $N_m$ in hand, $(N_{tot}-N_H-N_m)$ represents the number of largest particles; and, $N_m$ particles have the masses between $m_L$ and $m_H$; i.e., $m_i^{new} = m_H \left (\frac{m_L}{m_H} \right )^{\frac{i}{N_m+1}} $ with $i=1, 2, \dots, N_m$. Finally, the particles' smoothing lengths are updated proportionally to the new masses assigned, i.e., by the factor of $\sqrt{\frac{m_i^{new}}{m_i}}$ (in 2D simulations). 


\end{enumerate}

After the new masses and smoothing lengths of fluid particles are determined in the REFINE/COARSEN step, the distribution of particles is adapted using a particle-shifting technique. To this end, a shifting vector $\delta \textbf{r}_{i}$ is first defined as:
\begin{equation}
\delta \textbf{r}_{i} = \frac{\beta r_0^2 \Delta x_{ini}h_i}{{\bar m}_i h_{ini}}\sum\limits_j  m_j\frac{\textbf{r}_{ij}}{r_{ij}^3} \; ,
\label{equ:shifting_vector}
\end{equation}
where $r_0=\frac{1}{N_i}\sum\limits_j  {r_{ij}}$; $N_i = |{\cal{N}}_{h,i}|$; ${\bar m}_i = \sum\limits_j  m_j$; and $\beta$ is an adjustable dimensionless parameter and set as $\beta=0.05$ in this work. With this shifting vector, the position of particle $i$ is then update as:
\begin{equation}
\textbf{x}_{i}^{new} = \textbf{x}_{i} + \delta \textbf{r}_{i} \; .
\end{equation}
Based on its new position, each particle receives a small velocity and density correction via a second-order interpolation:
\begin{equation}\label{equ:correct_var}
\psi_{i}^{new} = \psi_{i} + \nabla \psi_{i}\cdot \delta \textbf{r}_{i} \; ,
\end{equation}
where the field variable $\psi_{i}$ represents $\textbf{v}_{i}$ or $\rho_{i}$,  and $\nabla \psi_{i}$ is calculated according to Eq.~\eqref{equ:gra_ope}. The second-order accuracy of this interpolation scheme matches that of the consistent SPH discretization. For consistency, the pressure $p_{i}^{new}$ is also corrected using $\rho_{i}^{new}$ via the EOS (Eq.~\eqref{equ:state-equ}). After the particle positions are shifted, the REFINE/COARSEN step concludes. A new distribution of variable resolutions is generated for the solution at the next time step. 

The proposed adaptive algorithm has three salient attributes. First, the adaptive-resolution implementation does not pose any additional constraints on $\Delta t$. Therefore, the computational efficiency gained only depends on the number of particles saved and the overhead involved in the adaptive algorithm. 

Second, this three-step procedure can be iterated for multiple times to attain an optimal distribution of adaptive resolutions. In the error-based criterion, we set a tolerance for the total relative error and terminate the iterations once the error satisfies:
\begin{equation}\label{equ:Error_TOL}
\frac{\sum_i \|\mathbf{e}_i\|^2}{\sum_i m_i \| \nabla \mathbf{v}_i \| ^2} \le TOL \; ,
\end{equation}
where $TOL$ is the preset tolerance. Through our numerical tests, we found that by and large one iteration per time step is sufficient to achieve the desired accuracy and savings of computational cost. 

Third, starting with a uniform resolution might not be optimal in terms of accuracy and efficiency, especially when transient dynamics is to be captured in simulating FSI with freely moving solid bodies. 
Starting with a uniform resolution takes much more iterations for adapting resolutions in each time step especially during the early time steps. As a result, the particle positions are shifted significantly across the streamlines, 
adversely impacting the accuracy of the solution. Therefore, we prefer to ``prime'' these simulations with a preprocessing step, in which the solid body is fixed at its initial position, and the fluid particles are adapted following the adaptive algorithm described above using the distance-based criterion. The adaptive process iterates until a desired distribution of variable resolutions is attained, in which the shifting vectors computed from Eq. \eqref{equ:shifting_vector} in a new iteration are considerably small; i.e., $\frac{1}{\beta N_{tot}} \sum_i \frac{|\delta \textbf{r}_{i}|}{\sqrt{m_i/\rho_i}} $ is less than a preset threshold. Provided this ``primed'' particle distribution, we commence the actual dynamic simulation. The three-step adaptive algorithm (error-based or distance-based) is subsequently applied at the end of each time step, yet the number of iterations and the magnitudes of the shifting vectors are greatly reduced. This ``prime-before-you-start'' approach improves both the accuracy and efficiency of the numerical solution.

\section{Simulation results}\label{sec:Simu_results}
Six different problems were studied to assess the accuracy, convergence, and efficiency of the proposed consistent adaptive-resolution SPH method. In this exercise, the numerical results were compared with analytical solutions, experimental data, or numerical solutions obtained using either the FEM or consistent SPH method with uniform, high resolutions. Unless otherwise specified, in all simulations the fluid had $\rho = 10^3~kg/m^3$ and kinetic viscosity $\nu = 10^{-4}~m^2/s$. 

\subsection{Transient Poiseuille flow}\label{subsec:Poiseuille_Flow}

First, we considered the transient Poiseuille flow in a straight channel with two fixed walls oriented along the $x$ axis and positioned at $y=0~m$ and $y=0.1454~m$. The flow was driven by a body force ($|\textbf{g}| = 2 \times 10^{-4}~m/s^2$) acting along the $x$ axis and subject to no-slip BCs at the two walls and periodic BCs at the remaining boundaries.  

\begin{figure}[H]
	\centering
	\includegraphics[width=5in]{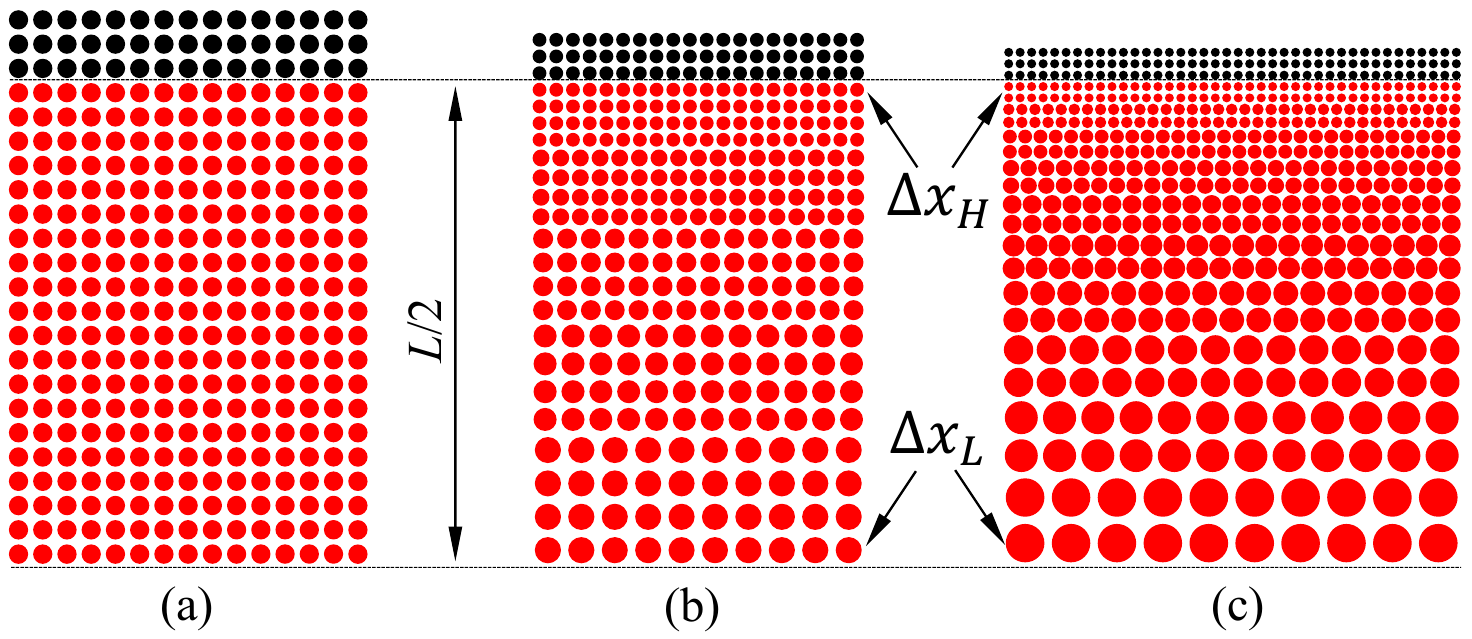} 
	\caption{Initial distribution of SPH particles with uniform/variable resolutions for modeling the transient Poiseuille flow. Due to the symmetry, only half of the simulation domain is shown. 
(a) Uniform resolution with $\Upsilon_{max} = $ 1; (b) adaptive resolution with $\Upsilon_{max} = $ 2; and, (c) adaptive resolution with $\Upsilon_{max} = $ 4.}
	\label{fig:Case1-Ini-Dis}
\end{figure}

\begin{figure}[H]
	\centering
	\includegraphics[width=4in]{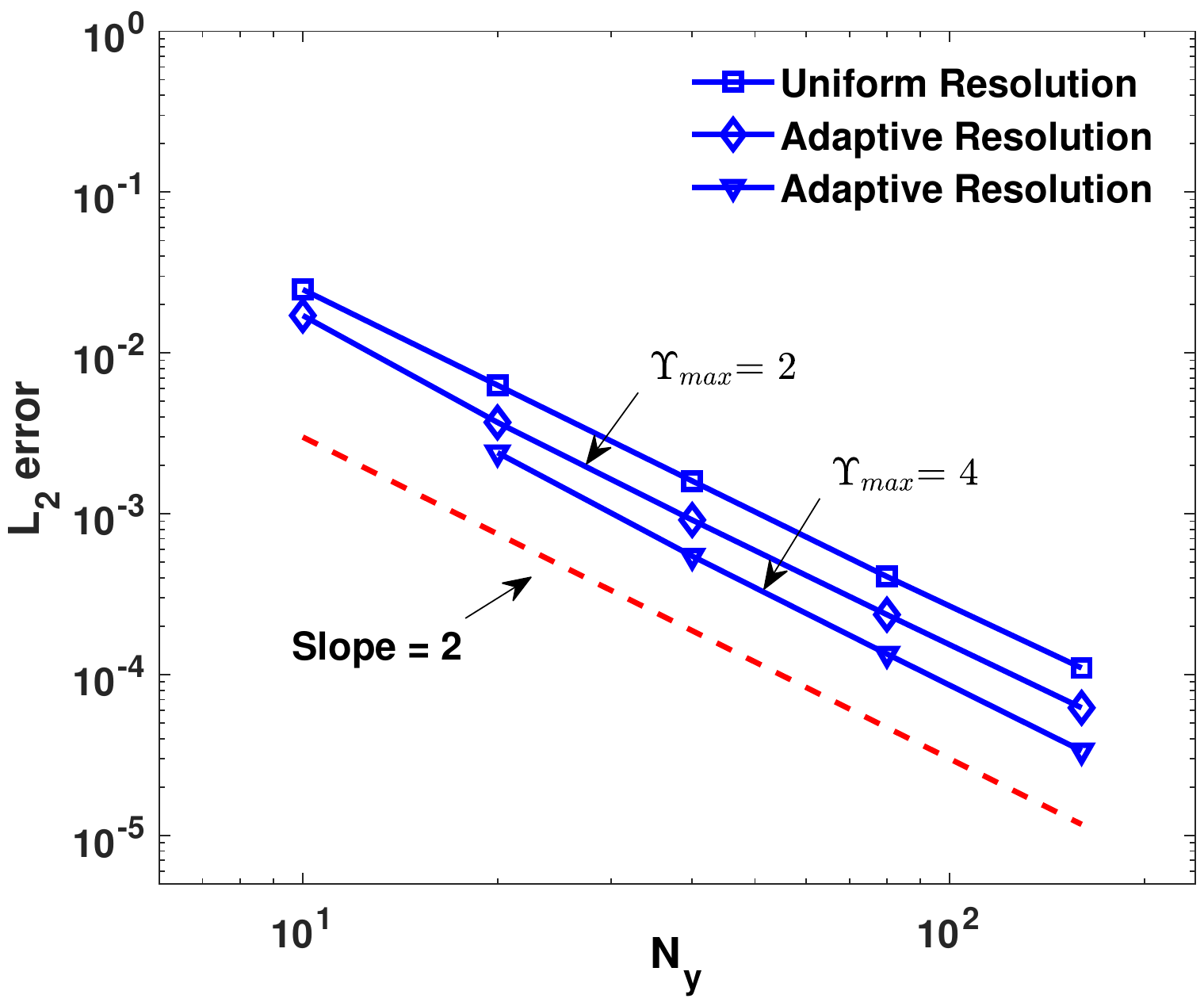} 
	\caption{Convergence of the numerical solutions on the steady-state velocity in Poiseuille flow for different $\Upsilon_{max}$. The $x$ axis displays the number of particles across the channel width; the $y$ axis is the $L_{2}$ error relative to the analytical solution.}
	\label{fig:Case1-Convergence}
\end{figure}

To assess the convergence of the adaptive-resolution numerical solution, we examined two different prescribed configurations of variable resolution. As shown in Figure \ref{fig:Case1-Ini-Dis}, the simulation domain was discretized by 5 or 10 different resolutions across the channel width with $\Upsilon_{max} = $ 2 or 4, respectively. In either case, the ratio of any two adjacent resolutions is less than 1.33. The coarsest resolution ($\Delta x_L$) is near the center of the channel; the finest resolution ($\Delta x_H$) is near the channel walls.  
For each of the three cases considered; i.e., $\Upsilon_{max}=1,\:2$, and $4$, we assessed the convergence of the SPH solutions by varying the number of particles and monitoring the errors. We report $L_2$ error by comparing the numerically predicted steady-state velocity with the analytical solution. The results are presented in Figure \ref{fig:Case1-Convergence}. First, the second-order convergence of the consistent adaptive-resolution SPH method can be obtained at all $\Upsilon_{max}$ values. Second, with the same number of particles across the channel width, larger $\Upsilon_{max}$ leads to smaller errors and hence higher accuracy, which implies that using variable resolution reduces the number of particles (or degrees of freedom) required to achieve a desired accuracy. 

\subsection{Flow around a periodic array of fixed cylinders}\label{subsec:Flow_Cylinder} 

\begin{figure}[H]
	\centering
	\includegraphics[width=6in]{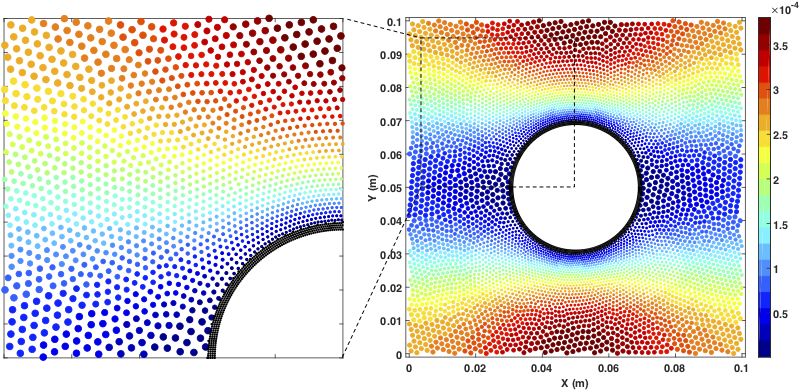}
	\caption{Distribution of SPH particles in the flow around a fixed cylinder at steady state with the resolutions adapted according to the recovery-based error estimator. The color is correlated to the flow velocity magnitude (unit in the color bar: $m/s$)}.
	\label{fig:Case2-Final-Conf}
\end{figure}

In this test, a cylinder of radius $0.02~m$ was positioned at the center of a square domain of length $0.1~m$. The flow was driven by a body force $|\textbf{g}| = 5 \times 10^{-5}~m/s^2$ along the $x$ coordinate and subject to the no-slip BC at the surface of the cylinder and periodic BCs at the remaining boundaries. We adapted the spatial resolutions dynamically using the recovery-based error estimator computed at each time step. Initially, the SPH particles were distributed with a uniform resolution $\Delta x_{ini}=1.25~mm$. We prescribed the finest and coarsest spacing of particles in the adapted resolutions as $\Delta x_H = \frac{1}{4}\Delta x_{ini}$ and $\Delta x_L = \frac{\sqrt{31}}{4}\Delta x_{ini}$, respectively;  i.e., $\Upsilon_{max}=\sqrt{31}$. The steps followed to adjust the resolution are as described in \S \ref{sec:Adapt_Algo}. Figure \ref{fig:Case2-Final-Conf} shows the distribution of adaptive resolutions at the steady state. (The transient dynamics is shown by the animation in the Supporting Information.)

\begin{figure}
	\centering
	\includegraphics[width=4in]{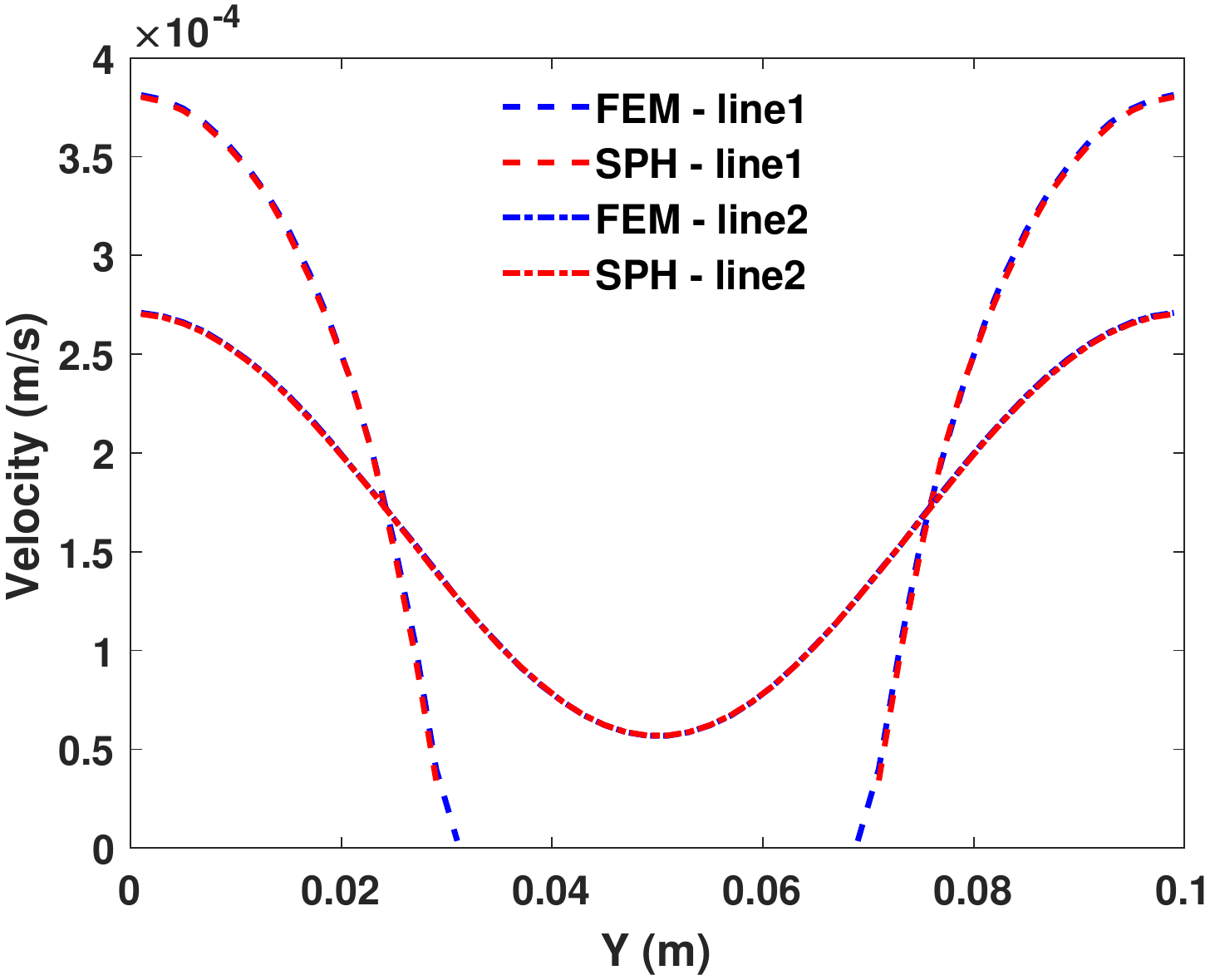}
	\caption{The steady-state velocity profiles along lines 1 ($x=0.05~m$) and 2 ($x=0.1~m$), computed by the consistent adaptive-resolution SPH with $\Delta x_H = 0.3125~mm$ and $\Upsilon_{max}=\sqrt{31}$ and FEM with a uniform resolution $\Delta x = 0.25~mm$.}
	\label{fig:Case2-FEM-SPH}
\end{figure}

To examine the accuracy of this numerical solution, the velocities along two vertical imaginary lines at $x=0.05~m$ (line 1) and at $x=0.1~m$ (line 2) were compared with those predicted by FEM with a higher, uniform resolution of $\Delta x = 0.25~mm < \Delta x_H$. As shown in Figure \ref{fig:Case2-FEM-SPH}, the SPH solutions agree well with the FEM solutions. We also compared the total drag force exerted on the cylinder by the fluid. The difference of our solution relative to that predicted by FEM is less than $1\%$. These results confirm the accuracy of the proposed method in modeling this flow. 

In addition to accuracy, we examined the computational efficiency of the adaptive-resolution SPH in modeling this flow. To that end, we also conducted an SPH simulation using the uniform, high resolution $\Delta x_H$. This led to a number of particles 16 times higher than that used in the adaptive-resolution simulation. The uniform-resolution computational cost (in terms of computer time) was about 8 times that of the adaptive-resolution simulation. By using the adaptive-resolution technique in this test, we saved about 93\% degrees of freedom and 87\% computer time without sacrificing accuracy.

\subsection{Flow around a cylinder in a narrow channel}\label{subsec:Flow_Channel}  
In this and the next tests, we simulated flows through narrow gaps between solid boundaries. The solid boundaries are either stationary or moving. The adaptive-resolution technique allowed for resolving the narrow gaps with sufficient resolutions but without increasing the total degrees of freedom of the simulations. 

\begin{figure}[H]
	\centering
	\includegraphics[width=4in]{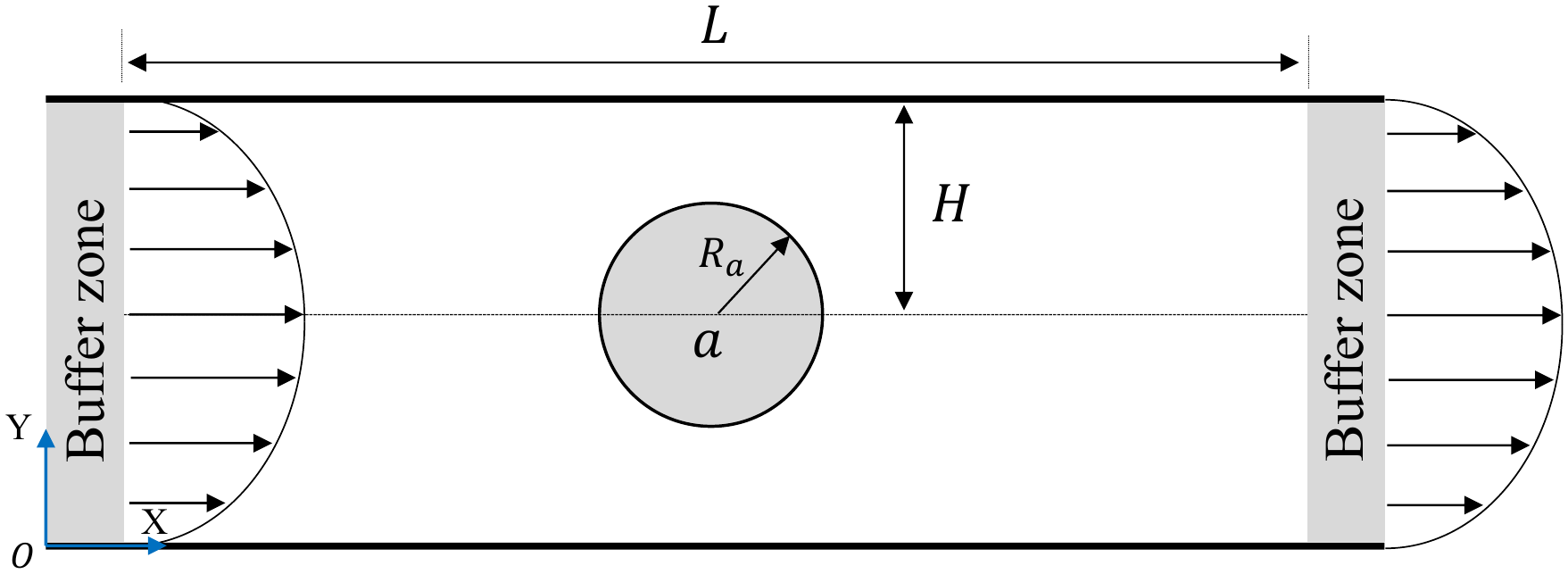}
	\caption{Configuration of the flow around a cylinder centered in a channel. Here, $L=0.3~m$; $2H=0.1~m$; and, $R_a$ varies from $0.1H$ to $0.9H$. Two buffer zones of ghost particles with a length of $0.02~m$ are set next to the inlet and outlet boundaries, respectively.}
	\label{fig:Case3-Ini-Conf}
\end{figure}

First, we consider a flow around a cylinder in a narrow channel. Figure \ref{fig:Case3-Ini-Conf} describes the setup of this problem. The computational domain was a box of length $L=0.3~m$ and width $2H=0.1~m$. A cylinder was at the center of the channel with the radius varied from $R_a = 0.1H$ to $R_a = 0.9H$. The larger the cylinder, the narrower the gap between the cylinder and channel walls. At the channel walls and surface of the cylinder, the flow was subject to no-slip BCs. To impose the prescribed inflow and outflow velocities at the inlet and outlet of the channel, two buffer zones of length $0.02~m$ were attached at the two ends. Particles in these buffer zones served as ghost particles for enforcing the Dirichlet boundary condition. The prescribed velocities were from the analytical solution of the steady-state Poiseuille flow. When particles moved out of the channel with flow, periodic boundary conditions were enforced. 

\begin{figure}[H]
	\centering
	\includegraphics[width=6in]{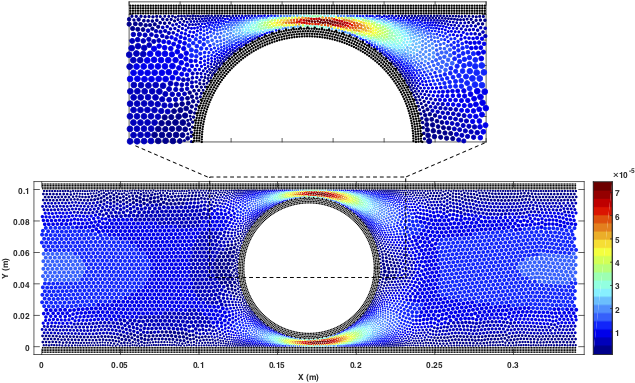}
	\caption{Distribution of SPH particles in the flow around a cylinder in a narrow channel at the steady state with the resolutions adapted according to the recovery-based error computed during the simulation. The color is correlated to the flow velocity magnitude (unit in the color bar: $m/s$).}
	\label{fig:Case3-Final-Conf}
\end{figure}

SPH particles were initially distributed with a uniform spacing $\Delta x_{ini}=2~mm$. For the cylinder of $R_a = 0.9H$, this resolution led to only 2 particles within the gap between the cylinder boundary and channel walls. For the kernel length used in the SPH discretization, at least 5 particles across the gap are needed to guarantee an accurate solution of the flow. However, this level of resolution throughout the simulation domain calls for prohibitively expansive simulations. Therefore, we maintained sufficiently fine resolutions within the gap but coarser resolutions for the remaining domain. Similarly to the previous case, we adapted the spatial resolutions according to the error estimator. We chose the finest resolution $\Delta x_H = \frac{1}{2}\Delta x_{ini}$ and the coarsest resolution $\Delta x_L = \frac{\sqrt{7}}{2}\Delta x_{ini}$, which gives $\Upsilon_{max} = \sqrt{7}$. Figure \ref{fig:Case3-Final-Conf} shows the distribution of adaptive resolutions at steady state. We note that the recovery-based error of velocity gradient effectively guided the redistribution of particles to the desired variable resolutions.

\begin{figure}
	\centering
	\begin{subfigure}[t]{0.47\textwidth}
	\centering
\includegraphics[height=2.5in]{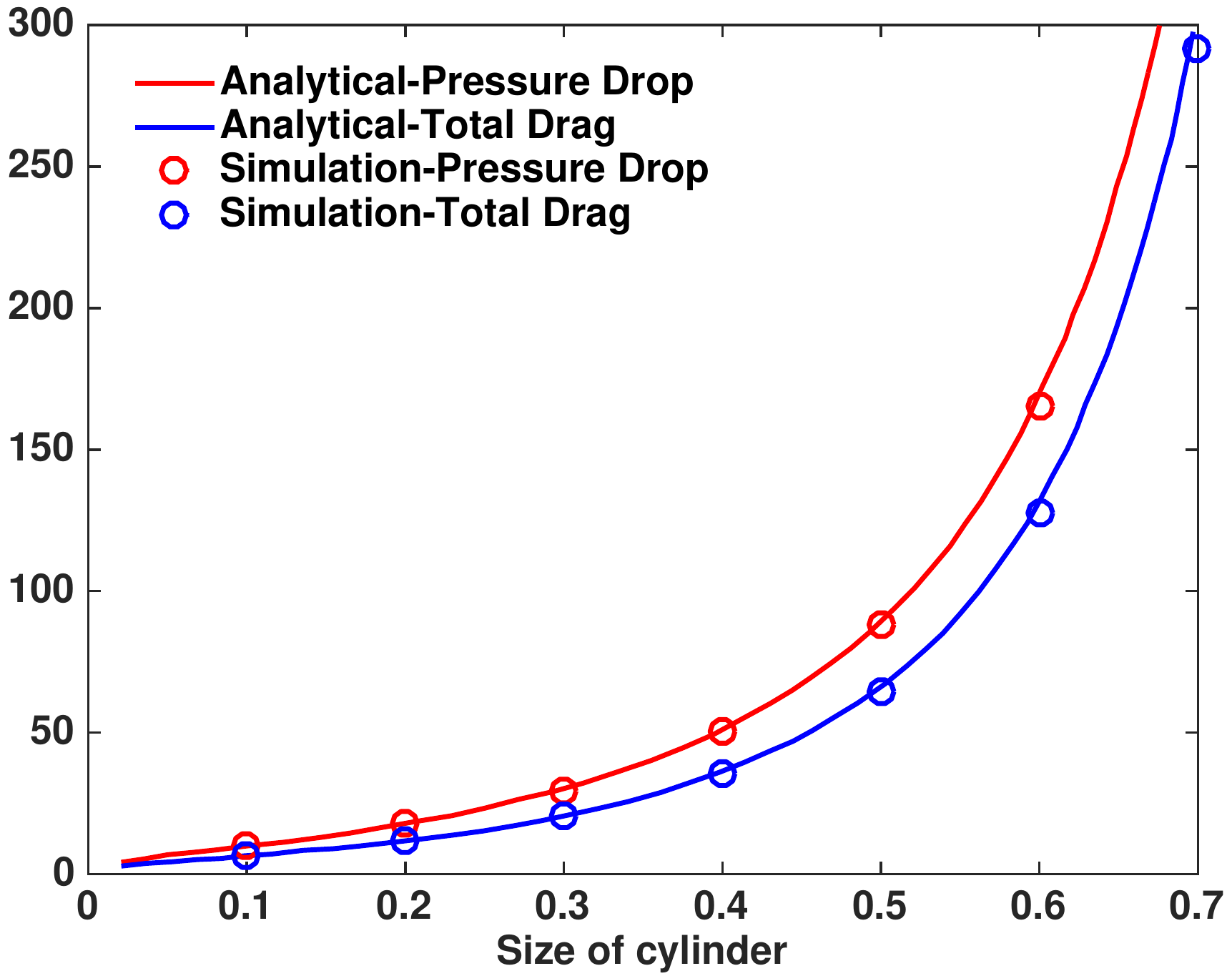}
	\caption{} \label{fig:Case3-Drag-force}
	\end{subfigure}
	\begin{subfigure}[t]{0.47\textwidth}
	\centering	\includegraphics[height=2.5in]{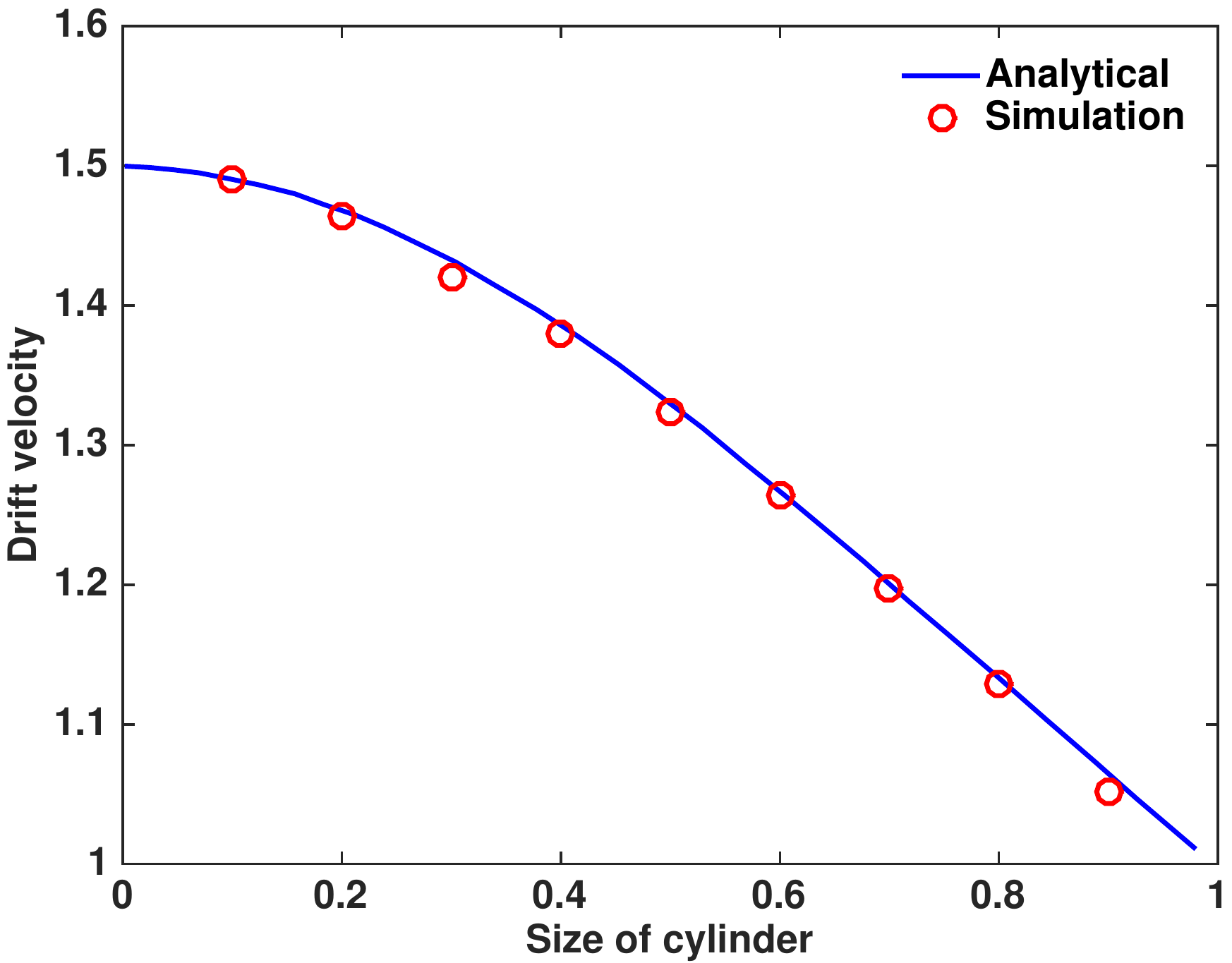}	
	\caption{} \label{fig:Case3-Drift-Velocity}
	\end{subfigure}
	\caption{(a) The total drag force exerted on the cylinder and pressure drop of the flow. (b) The drift velocity of the cylinder. Data are collected for different sizes of the cylinder (defined as $R_a/H$). Numerical results are compared with the analytical solutions \cite{jeong2014two}.}
\end{figure}

We monitored the drag force exerted on the cylinder as well as the pressure drop of the flow through the channel and compared them with analytical solutions, see Figure \ref{fig:Case3-Drag-force}.
As the cylinder radius ($\frac{R_a}{H}$) increases, the total drag force ($\frac{F}{2\rho\nu V_r}$) exerted on the cylinder grows rapidly and so does the pressure drop ($\frac{\Delta P H}{\rho\nu V_r}$) of the flow, where $V_r$ is the mean velocity of the prescribed Poiseuille flow. Both the drag and pressure drop solutions agree well with their analytical counterparts. To further assess the solution accuracy, we conducted an additional set of simulations, from which we computed the drift velocity 
of the cylinder driven by the flow. Following the work by Jeong and Yoon \cite{jeong2014two}, in this set of simulations the flow was set static initially, but the cylinder was moving at constant velocity equal to $V_r$. Again, we computed the drag force exerted on the cylinder by the fluid. The drift velocity is equal to the ratio of the drag force computed from the previous set of simulations to the drag force computed from this set of simulations \cite{jeong2014two}.  Figure \ref{fig:Case3-Drift-Velocity} compares the drift velocity thus obtained with the analytical solution \cite{jeong2014two}, and again, a good agreement is found. We hence demonstrated the accuracy of the proposed adaptive-resolution SPH method in simulating a flow with narrow gaps between solid boundaries.

\subsection{Dynamics of two cylinders under a shear flow}\label{subsec:Cylinder_Shear_Flow} 

\begin{figure}[H]
	\centering
	\includegraphics[width=3in]{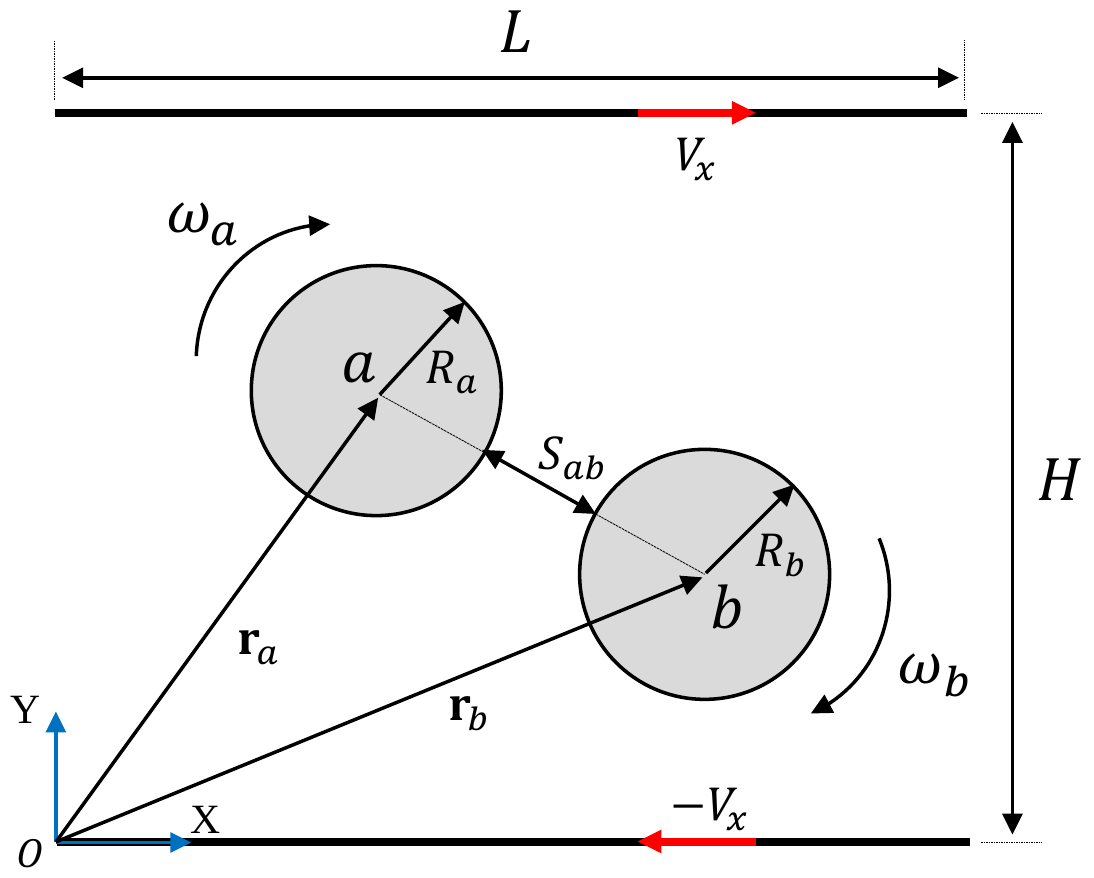}
	\caption{Configuration of two cylinders under a shear flow.}
	\label{fig:Case4-Ini-Conf}
\end{figure}

Next, we simulated the dynamics of two cylinders immersed in fluid subject to a plane shear flow. Figure \ref{fig:Case4-Ini-Conf} illustrates the setup of this problem. The radius of the equal-sized cylinders is $R_a=R_b=0.01~m$. Periodic BCs were applied at $x=0$ and $x=0.4~m$, and the solid walls were placed at $y=0$ and $y=0.4~m$ with the no-slip BC. Thus, the size of the simulation domain is $L=H=0.4~m$, which is large enough to neglect the effects of walls and periodic images of cylinders. To generate a plane shear flow with the shear rate of $\dot{\gamma}=\frac{2V_x}{H}=1.25\times10^{-3}/s$, the upper and lower walls were assigned velocities in opposite directions but with the same magnitude of $V_x=2.5\times10^{-4}m/s$. In this case, the kinetic viscosity of the fluid was assumed as $\nu = 10^{-5} m^2/s$. With that, the Reynolds number defined as $Re=\frac{\dot{\gamma}{R_a}^2}{\nu}$ has the same value as that in the literature \cite{bian2014splitting,darabaner1967particle,trask2017compatible}, i.e., $Re=0.0125$. Following the flow, the two cylinders rotated while approaching each other or moving apart. 

In literature, this problem was previously solved both analytically \cite{darabaner1967particle} and numerically \cite{bian2014splitting,trask2017compatible}. Varying the initial positions of the two cylinders, the translational trajectories of either cylinder's center of mass can be different. 
Following \cite{bian2014splitting,trask2017compatible}, we also considered three different initial positions of the cylinders. Correspondingly, the initial separations of the two cylinders in $(x, y)$ were: $(0.03~m, 0.02~m)$, $(0.03~m, 0.012~m)$, and $(0.024~m, 0.0~m)$, respectively. For a fair comparison to the analytical solution derived by assuming a Stokes flow and neglecting the inertia effect \cite{darabaner1967particle}, we assumed in our simulations that the cylinders were rotating with the angular velocity specified in the analytical solution. 

In this problem, the fluid occupies most of the domain and is subject to shear flow with a constant shear rate. As a result, the recovery-based errors of velocity gradient showed very little difference across the entire domain. Thus, the error-based adaptive criterion was not effective here. Alternatively, we used the distance to the surfaces of the two cylinders as the criterion for adapting the spatial resolutions in this test in which we attempted to capture transient dynamics of freely moving cylinders. As discussed in \S \ref{sec:Adapt_Algo}, starting the simulation with a uniform resolution was not optimal in terms of accuracy and efficiency of the solution. Instead, we carried out a preprocessing step before the simulation. Therein, the two cylinders were fixed at their initial positions and the spatial resolution was adapted to a desired distribution using the distance-based adaptive algorithm. To that end, we started with a uniform resolution $\Delta x_{ini}=4~mm$ and set $\Delta x_H=\frac{1}{4}\Delta x_{ini}$, $\Upsilon_{max}=\sqrt{31}$ (or $\Delta x_{L} = \frac{\sqrt{31}}{4} \Delta x_{ini}$), and $N_H = 400$ in Eq. \eqref{equ:preadapt_Nm}. The adaptive algorithm was executed iteratively until the shifting vectors computed from Eq. \eqref{equ:shifting_vector} in a new iteration were sufficiently small; i.e., $\frac{1}{\beta N_{tot}} \sum_i \frac{|\delta \textbf{r}_{i}|}{\sqrt{m_i/\rho_i}} < 6\times 10^{-3}$. 

\begin{figure}[H]
	\centering
	\includegraphics[width=6in]{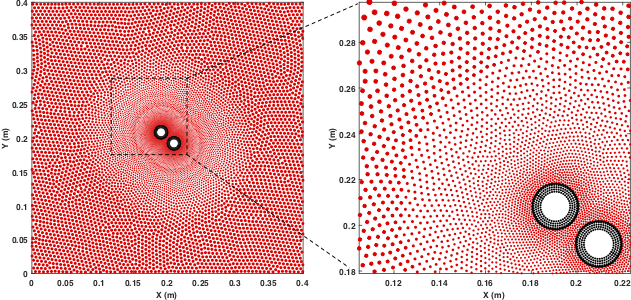}
	\caption{A snapshot of the distribution of SPH particles around the two cylinders under a shear flow with the resolutions adapted according to the distances to the cylinders ($t = 1500~s$).}
	\label{fig:Case4-Final-Conf}
\end{figure}

The actual simulation was ``seeded'' with the variable resolutions generated from the preprocessing step. We continued to use the distance-based adaptive algorithm while the two cylinders and fluid particles were moving with the flow. Figure \ref{fig:Case4-Final-Conf} presents a snapshot of the distribution of SPH particles with the adapted resolutions, in which the two cylinders had the initial separation of $(0.03~m,~0.012~m)$. (The entire dynamics is shown by the animation in the Supporting Information.) As seen in Figure \ref{fig:Case4-Final-Conf}, the resolutions were gradually coarsened away from the cylinders. With the finest resolution near the cylinders, we maintained sufficient particles to resolve the gap between the two cylinders when they approached each other closely. 

\begin{figure}
	\centering
	\includegraphics[width=4in]{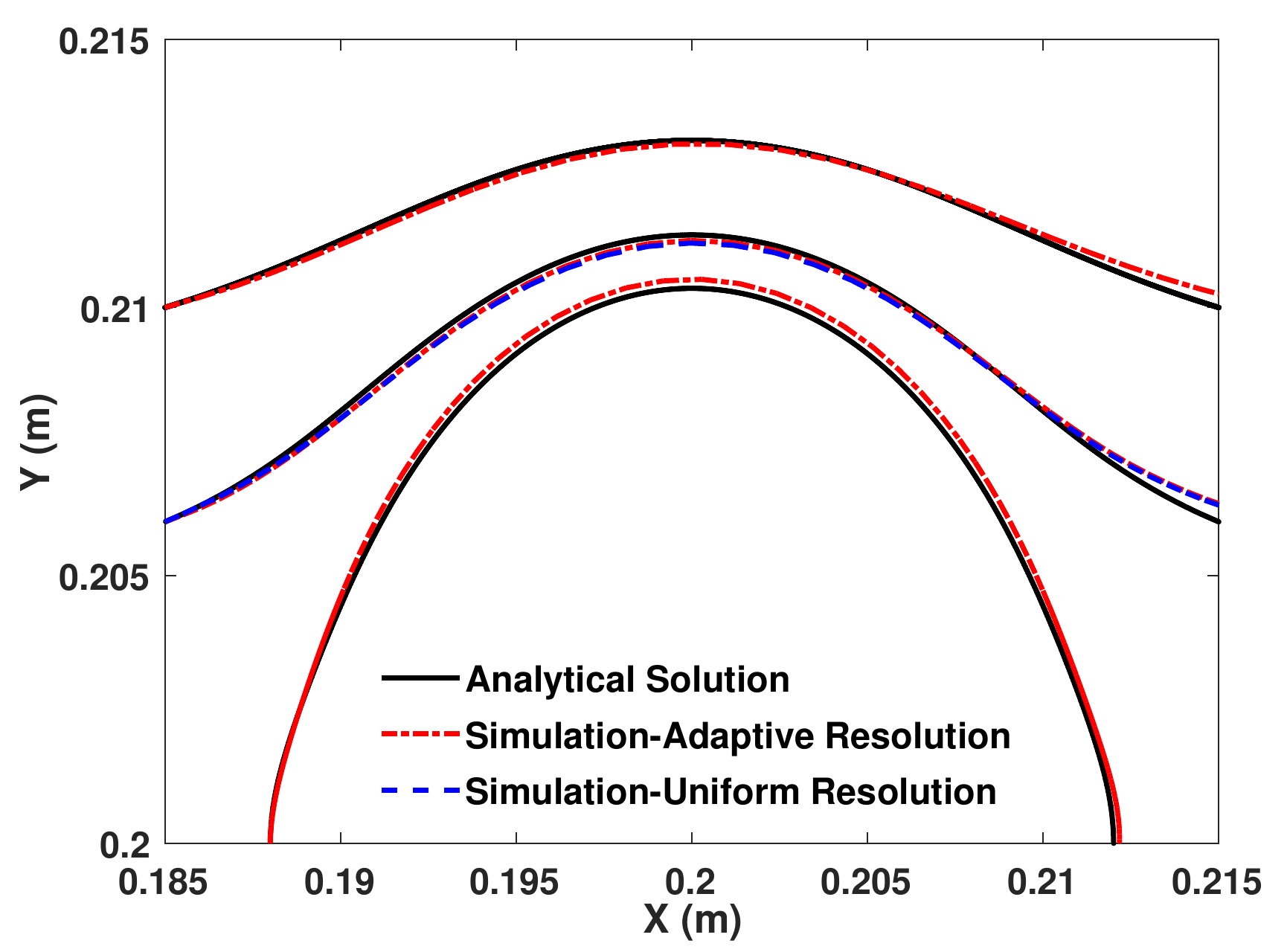}
	\caption{Trajectories of one of the two cylinders under the shear flow with three different initial positions, computed from the consistent SPH simulations and compared with the analytical solutions \cite{darabaner1967particle}.}
	\label{fig:Case4-Trajectory}
\end{figure}

Due to the symmetry of the two cylinders, we only tracked the trajectory of the left cylinder's center of mass. Figure \ref{fig:Case4-Trajectory} depicts the three trajectories computed from our simulations and compared with the analytical solutions \cite{darabaner1967particle}. Results produced by the consistent adaptive-resolution SPH satisfactorily agree with the analytical solutions. We also provide in Figure \ref{fig:Case4-Trajectory} the trajectory predicted using the uniform high resolution of $\Delta x=\Delta x_H=1~mm$, which overlaps the curve obtained from the adaptive-resolution simulation. In the case of the bottom trajectory in Figure \ref{fig:Case4-Trajectory}, the two cylinders could get as close as less than one particle spacing ($\Delta x_H=1~mm$) between their boundaries. To avoid the singularity caused by lack of discretization points within the gap between the two cylinders, we followed Bian et al. \cite{bian2014splitting} and applied a repulsive interaction between the cylinders based on a lubrication model \cite{yuan1994rheology,kromkamp2006lattice} for both adaptive-resolution and uniform-resolution simulations. However, it was applied only when the two cylinders came within $\Delta x_H$ between each other. Note that the number of particles used in the adaptive resolution simulation was only $\frac{1}{16}$ of that in the previous work by Bian et al, who drew on an inconsistent SPH algorithm \cite{bian2014splitting}. Moreover, owing to the consistent nature of the discretization, our results show improved accuracy with better agreement with the analytical solutions. In terms of computational efficiency, the adaptive-resolution simulation saved about 87\% computer time (85\% if the pre-processing step was included) compared against the simulation with the uniform high resolution. 

\subsection{Dynamics of a L-shaped active colloid}\label{subsec:Lshape_colloid} 

\begin{figure}[H]
	\centering
	\includegraphics[width=3in]{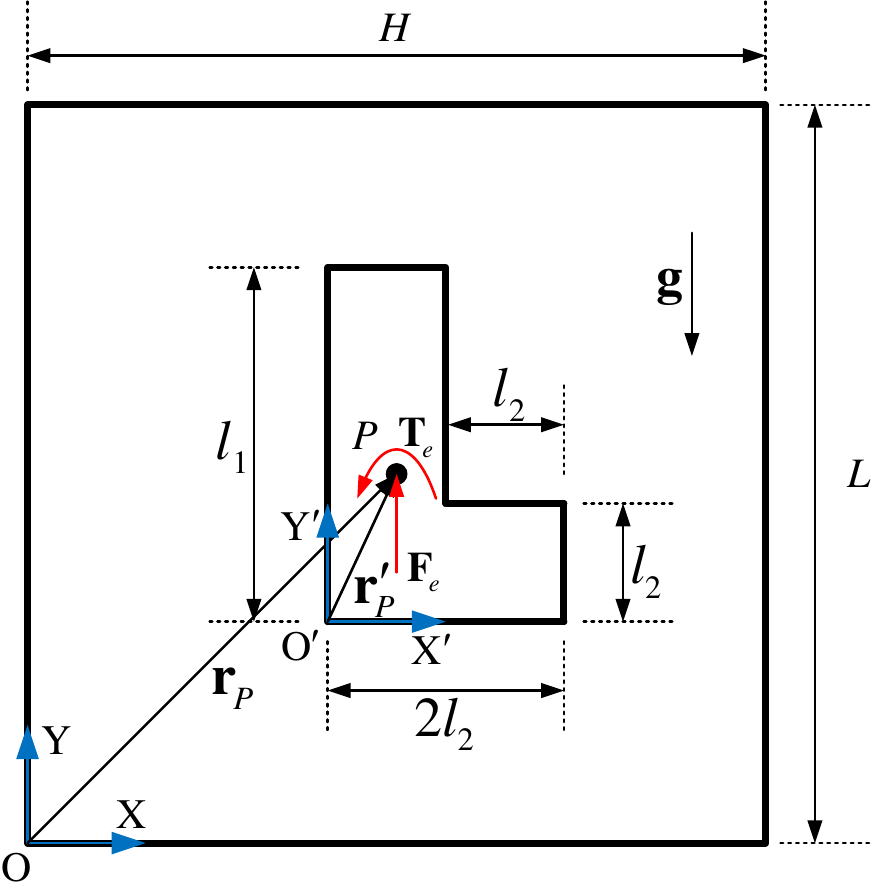}	
	\caption{Configuration of the simulation domain and L-shaped colloid. Here, $H = L = 100 \times 10^{-6} m$; $l_{1} = 18 \times 10^{-6} m$; and, $l_{2} = 6 \times 10^{-6} m$. The coordinates of the center of mass of the colloid are $(x'_p, y'_p) = (4.5 \times 10^{-6} m, 7.5 \times 10^{-6} m)$, with the origin $O'$ of the local coordinate system set at the bottom left corner of the ``L" shape. $\textbf{F}_e$ and $\textbf{T}_e$ represent the effective force and torque acting on the colloid to induce its self-propulsion.} \label{fig:Lshape-Geometry}
\end{figure}

In this section, we reproduced an interesting phenomenon concerning the motion of an asymmetric self-propelled colloid in fluid, which was first observed in experiments \cite{ten2014gravitaxis}. The colloid is L-shaped with two asymmetric arms. The colloid's motion was constrained to 2D in the experiment \cite{ten2014gravitaxis}. Thus, we set up a 2D simulation with a domain comparable to the chamber used in the experiment \cite{ten2014gravitaxis}, as depicted in Figure \ref{fig:Lshape-Geometry}. The no-slip BC is imposed on the surface of the L-shaped colloid; and, periodic BCs are applied at the remaining boundaries. The kinetic viscosity of the fluid was $\nu = 2.4 \times 10^{-9} m^2/s$. Both the colloid and fluid are subject to the gravitational force $\mathbf{g}$ {with $|\mathbf{g}|= 0.3~m/s^{2}$}. To mimic the colloid's self-propelling mechanism as in the experiment, an effective force $\textbf{F}_e$ and torque $\textbf{T}_e$ were applied at the center of mass of the L-shaped colloid. During the motion of the colloid, the magnitude of $\textbf{F}_e$ was kept constant, and it's always along the longer arm of the ``L" shape regardless of the orientation of the colloid. The torque $\textbf{T}_e$ is related to $\textbf{F}_e$ by $\textbf{T}_e = (l_2 - x'_p) \textbf{e}_{\textbf{x}'} \times \textbf{F}_e$, where $x'_p$ and $\textbf{e}_{\textbf{x}'}$ are defined in the local coordinate system on the colloid. By varying the magnitude of $\textbf{F}_e$, the L-shaped active colloid could display rich dynamic behavior owing to its nonlinear hydrodynamic interaction with the fluid.

\begin{figure}[htbp]
\centering
\begin{subfigure}[t]{1.0\textwidth}
\centering
\includegraphics[width=6in]					{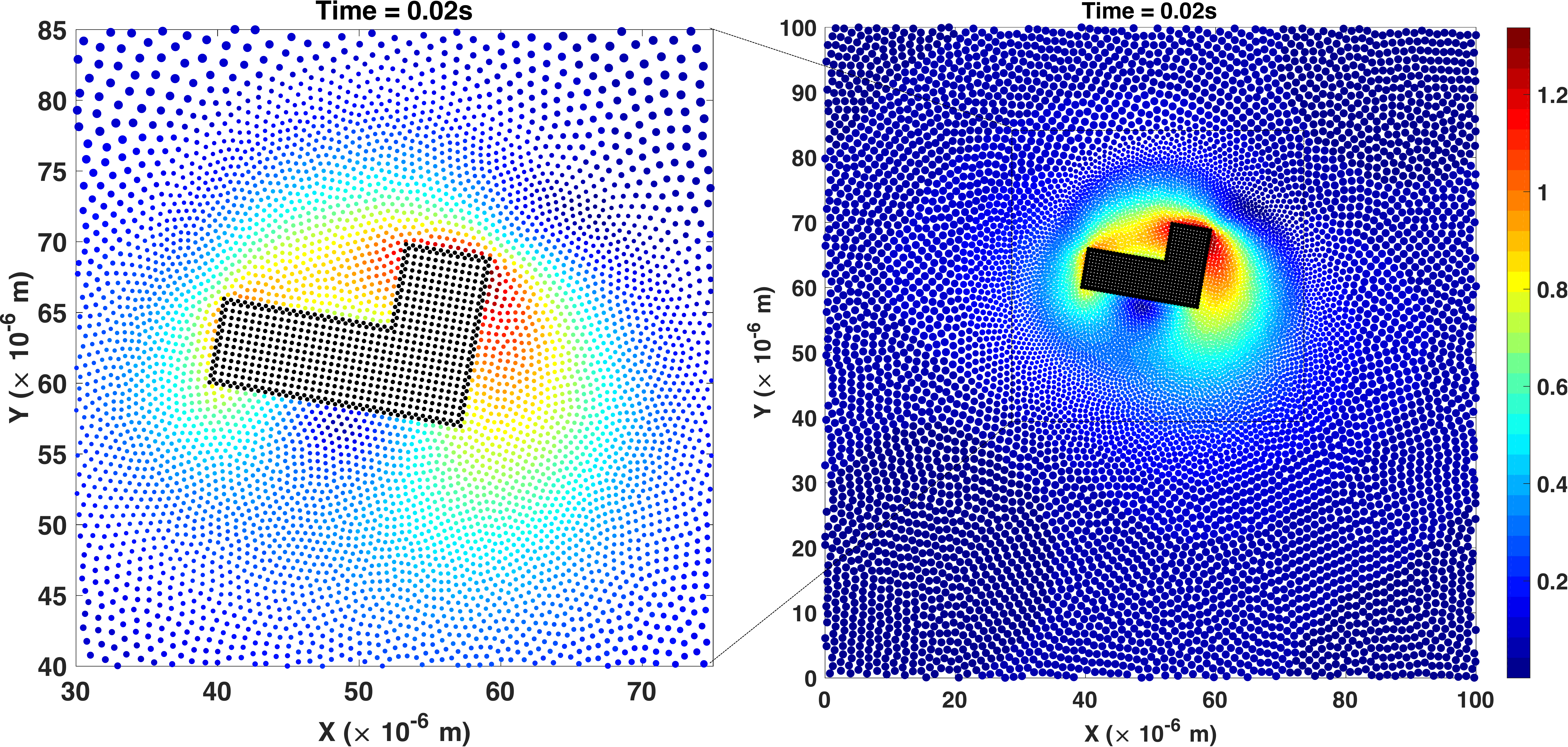}
\caption{$t = 0.02 s$} \label{fig:Lshape-velocity-2}
\end{subfigure}
\\
\begin{subfigure}[t]{1.0\textwidth}
\centering
\includegraphics[width=6in]					{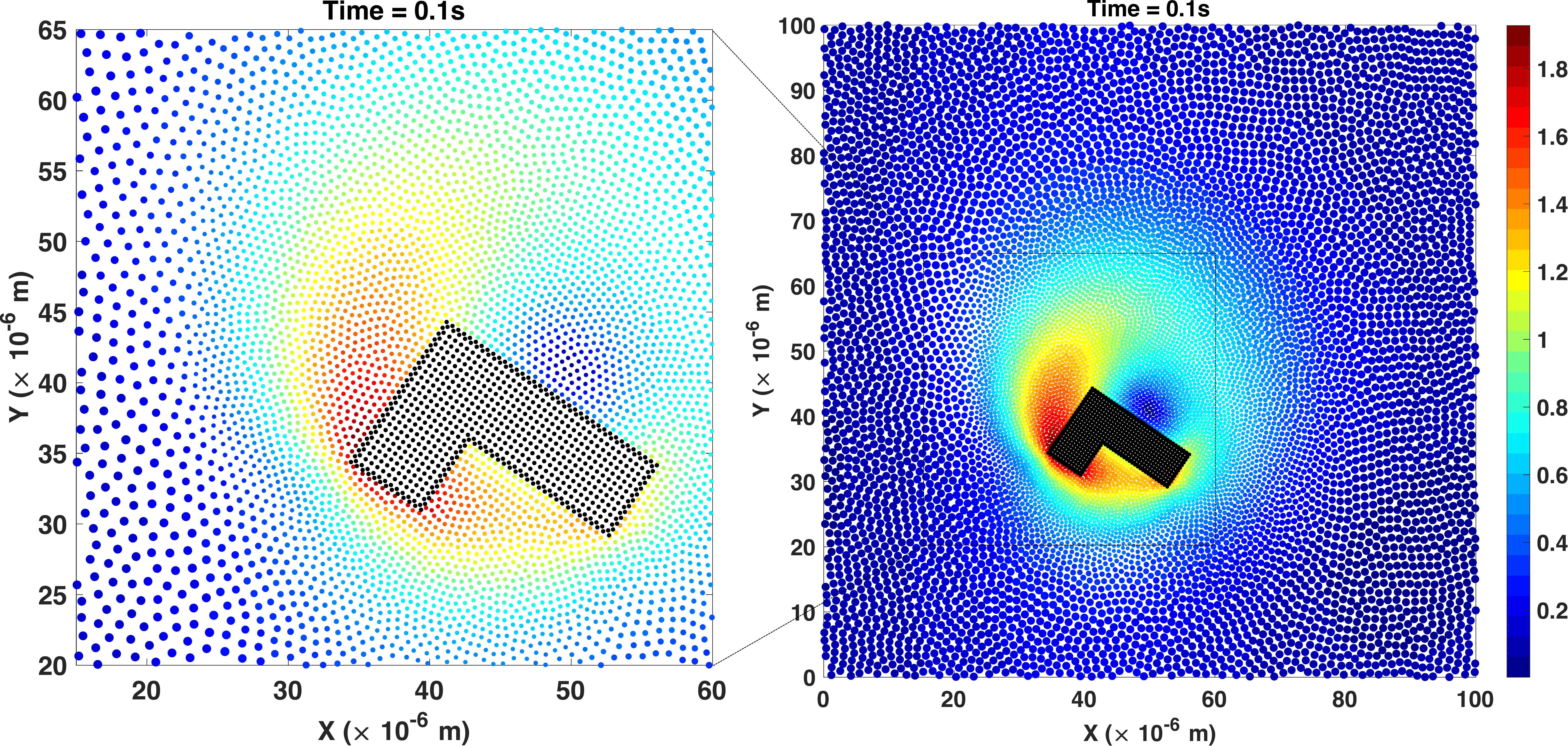}
\caption{$t = 0.1 s$} \label{fig:Lshape-velocity}
\end{subfigure}
\caption{Two snapshots of the distribution of SPH particles around the L-shaped colloid at $t = 0.02, 0.1 s$. The color is correlated to the flow velocity magnitude (unit in the color bar: $10^{-3} m/s$).}
\label{fig:Lshape-velocity-comparison}
\end{figure}

As in \S \ref{subsec:Cylinder_Shear_Flow}, a preprocessing step preceded the actual simulation. For preprocessing, the colloid was fixed at its initial position and the spatial resolution was adapted to a desired distribution using the distance-based adaptive algorithm. To that end, we started with a uniform resolution $\Delta x_{ini} = 1 \times 10^{-6}~m$ and set $\Delta x_{H} = \frac{1}{2} \Delta x_{ini}$, $\Delta x_{L} = \frac{\sqrt[]{7}}{2} \Delta x_{ini}$ (i.e., $\Upsilon_{max} = \sqrt[]{7}$), and $N_H = 700$. Iterations of the adaptive algorithm ended when the shifting vectors computed from Eq. \eqref{equ:shifting_vector} satisfied $\frac{1}{\beta N_{tot}} \sum_i \frac{|\delta \textbf{r}_{i}|}{\sqrt{m_i/\rho_i}} < 6 \times 10^{-3}$. The actual experiment commenced after this preprocessing step. During the simulation, the error-based adaptive algorithm was employed to dynamically adapt spatial resolutions while the colloid and fluid particles were moving with the flow. Figure \ref{fig:Lshape-velocity-comparison} presents two snapshots of the simulation, showing the distributions of adaptive resolutions around the L-shaped colloid and throughout the domain at two different time instances. (The entire dynamics is shown by the animation in the Supporting Information.)  

\begin{figure}[H]
	\centering
	\includegraphics[width=4in]{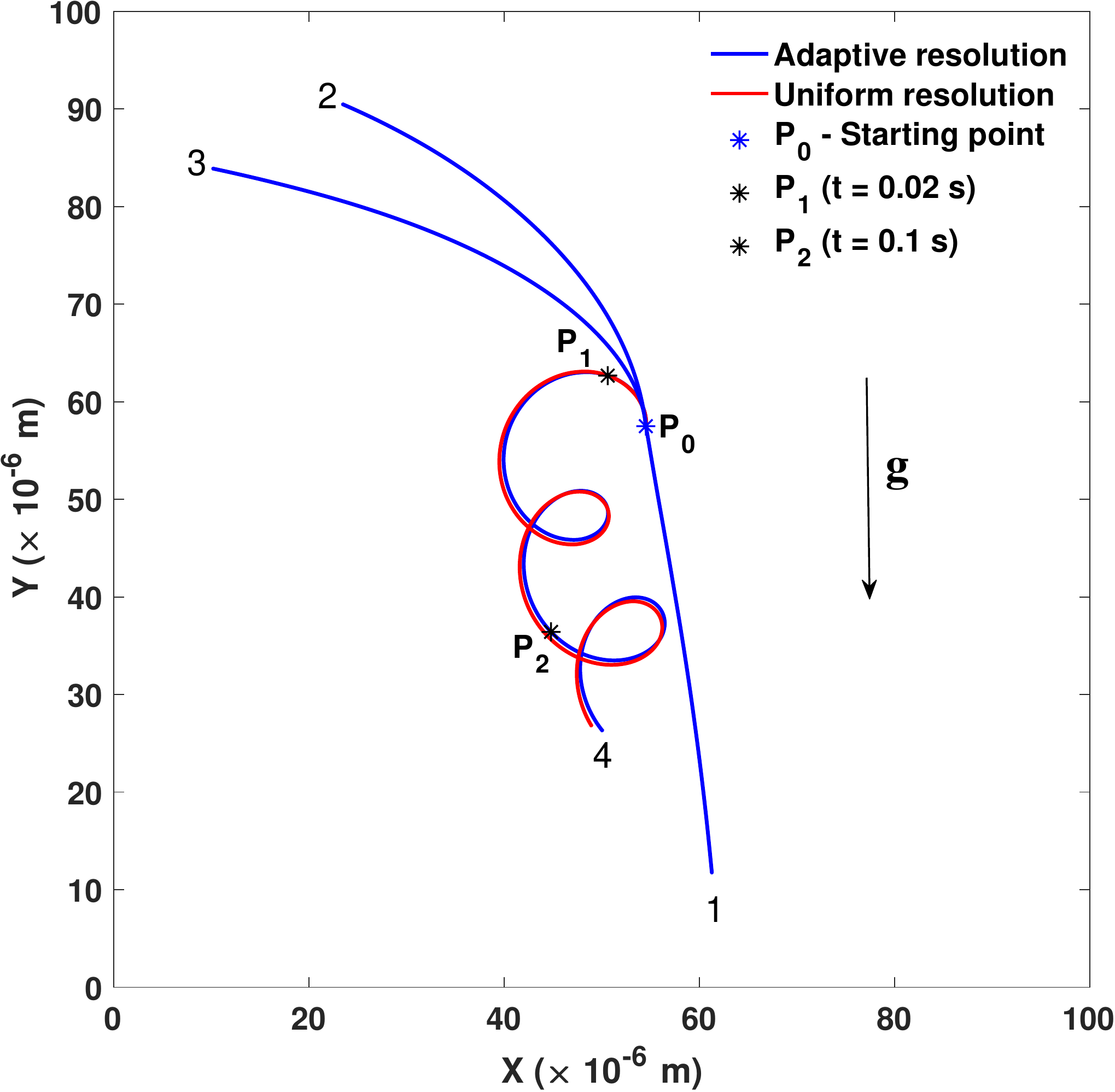}
	\caption{Trajectories of the L-shaped colloid at four different magnitudes of $\textbf{F}_e$. All trajectories start at the same position (blue asterisk). Trajectory 1 corresponds to the SDS with {$|\textbf{F}_e|=0.06~m/s^{2}$}; trajectory 2 and 3 correspond to the SUS with {$|\textbf{F}_e|=0.45, 0.6~m/s^{2}$}, respectively; and, trajectory 4 corresponds to the TLM with {$|\textbf{F}_e|=1.2~m/s^{2}$}. The two marked points (black asterisk) correspond to the two snapshots shown in Figure \ref{fig:Lshape-velocity-comparison} at $t = 0.02, 0.1 s$, respectively.}
	\label{fig:Lshape-Trajectory}
\end{figure}

As observed in the experiment \cite{ten2014gravitaxis}, with the magnitude of $\textbf{F}_e$ varied, the L-shaped active colloid followed different trajectories. For a small $\textbf{F}_e$, the colloid's self-propelling motion was slow. As a result, the impact of gravitational force was dominant, and hence, the colloid did a straight downward swimming (SDS). As $\textbf{F}_e$ gradually increased, the active motion of the colloid sped up, and its hydrodynamic interaction with the fluid started to compete with the impact of gravitational force and produced a trochoid-like motion (TLM) or even straight upward swimming (SUS).  
Our simulations were able to predict these three motions of the colloid. As illustrated in Figure \ref{fig:Lshape-Trajectory}, the trajectories of the colloid at four different magnitudes of $\textbf{F}_e$ display its SDS, TLM, and SUS motions. For the most complicated TLM, we further compared the trochoid-like trajectory predicted by the adaptive-resolution simulation with that computed from the uniform high-resolution ($\Delta x = \Delta x_{H} = 0.5 \times 10^{-6}~m $) simulation. A good agreement is found in the two solutions. Meanwhile, the adaptive-resolution simulation only used $\frac{1}{4}$ of SPH particles needed in the uniform high-resolution simulation and was two times faster in terms of computer time than the uniform high-resolution simulation.

\subsection{Collective dynamics of active colloids}\label{subsec:Mov_Col_Shear} 

Lastly, we simulated a more challenging problem -- the collective dynamics of a suspension of colloids. The colloids were actively rotating with a constant speed. As discussed in the literature \cite{goto2015purely}, these active colloids were driven by hydrodynamic interactions to repel or attract each other and hence could display rich collective behaviors. The simulation domain is $0.36~m$ by $0.24\sqrt{3}~m$. Initially, 48 equal-sized circular disks with a radius of $R_{d} = 0.0125~m$ were randomly placed in the domain. Each disk rotated clockwise with the same prescribed angular velocity of $\omega_{d} = 10^{-2} \:rad/s$. The kinetic viscosity of the fluid was $\nu =0.263\times 10^{-6} m^2/s$. This problem setup yields an area fraction of 0.04 for the colloid suspension and a Reynolds number (defined as $Re=\frac{\omega_{d} R^2_{d}}{\nu}$) of 5.94, which match those considered in the literature \cite{goto2015purely}. The flow was subject to no-slip BC at the surfaces of colloids and periodic BCs at the remaining boundaries. The spatial resolutions were adapted according to the recovery-based error estimator. Initially, the SPH fluid particles were distributed with a uniform spacing $\Delta x_{ini}= 4~mm$. The finest resolution was prescribed as $\Delta x_H=\frac{1}{2}\Delta x_{ini}$; and, the maximum resolution ratio was $\Upsilon_{max}=\sqrt{7}$. 

\begin{figure}[H]
	\centering
	\includegraphics[width=6in]{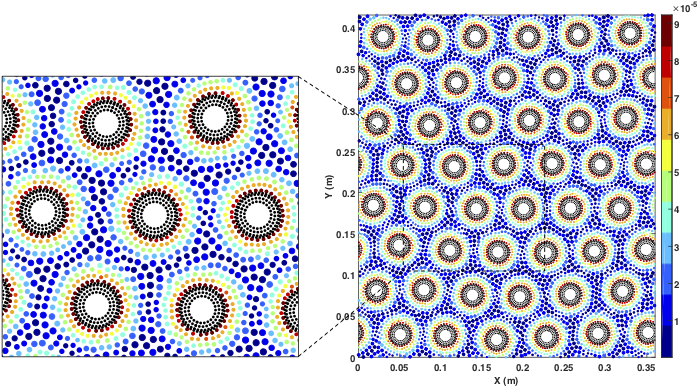}
	\caption{A suspension of active colloids with a hexagonal-ordered configuration at $Re=5.94$. The spatial resolutions were adapted according to the recovery-based error computed during the simulation. The color is correlated to the flow velocity magnitude (unit in the color bar: $m/s$).}
	\label{fig:Case5-Par-Dis-Adaptive}
\end{figure}

\begin{figure}[H]
	\centering
	\includegraphics[width=6in]{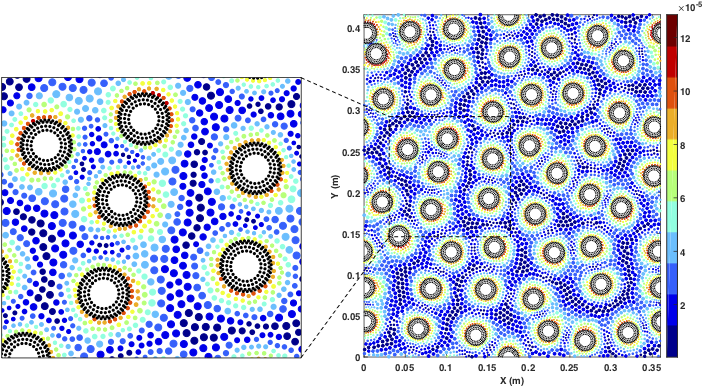}
	\caption{A suspension of active colloids randomly distributed at $Re=15.625$. The spatial resolutions were adapted according to the recovery-based error computed during the simulation. The color is correlated to the flow velocity magnitude (unit in the color bar: $m/s$).}
	\label{fig:Case5-Par-Dis-Adaptive-Re156}
\end{figure}

According to the work by Goto \cite{goto2015purely}, the collective behavior of these active disks depend on the Reynolds number. At $Re=5.94$, the colloids finally formed a stable hexagonal, ordered crystal-like configuration. We reproduced this observation in our simulation. Figure \ref{fig:Case5-Par-Dis-Adaptive} presents a snapshot of the colloids in this simulation, from which the hexagonal, ordered configuration can be clearly observed. Once this ordered configuration was formed, it remained stable throughout the entire simulation. When $Re$ was increased to 15.625, the system became unstable. The colloids moved around each other but could not form any ordered configuration, same as reported by Goto \cite{goto2015purely}. Figure \ref{fig:Case5-Par-Dis-Adaptive-Re156} shows a snapshot from the simulation with $Re=15.625$, in which the colloids are randomly distributed even after a long time simulation. In Figures \ref{fig:Case5-Par-Dis-Adaptive} and \ref{fig:Case5-Par-Dis-Adaptive-Re156}, we note that the error-based criterion gave rise to finer resolutions near the colloids' boundaries, which led to more accurate solutions for the near-field hydrodynamic interactions between colloids.   

\section{Conclusion}\label{sec:conclusion}

We have presented a new consistent adaptive-resolution SPH method. It accurately captures the near-field hydrodynamic variables in FSI with sufficiently high resolutions and enlarges the computational capability of FSI problems without increasing the computational cost. 

The method is based on the second-order consistent SPH discretization that ensures the accuracy of the numerical solution even when SPH particles of different sizes (variable resolutions) are coupled within a kernel support. Moreover, a three-step adaptive algorithm was proposed to dynamically adapt spatial resolutions during the simulation. A key element of this algorithm is the criterion of adaptivity, of which we assessed two. The first one is distance-based, in which the spatial resolutions in the fluid domain gradually vary according to the distances to the solid boundary. The second criterion is error-based and adapts the resolutions according to a \textit{posteriori} recovery-based error estimator and following the error equidistribution strategy. In practice, these two criteria can be combined to optimize the accuracy and efficiency of the numerical simulations. As demonstrated in one of the tests (\S \ref{subsec:Lshape_colloid}), the distance-based criterion was employed in preprocessing to generate an initial distribution of adaptive resolutions. The error-based criterion was subsequently used in the simulation for dynamically adapting resolutions. 

A noteworthy component of the adaptive algorithm is its particle-shifting technique employed for refining or coarsening resolutions. In the proposed approach, the SPH particles marked for refinement/coarsening had their masses changed first. Thereafter, shifting vectors were computed for positioning the particles at their new locations. Subsequently, the hydrodynamic variables were corrected using second-order interpolation to account for their new positions.

We applied the new method in simulating six different flow and FSI problems: the transient Poiseuille flow, flow around a periodic array of cylinders, flow around a cylinder in a narrow channel, two colloids rotating under a shear flow, dynamics of a L-shaped active colloid, and collective dynamics of a suspension of active colloids. The numerical results were compared with analytical predictions, experimental data, and simulation results from FEM or the consistent SPH with a uniform, high resolution. For all tests we noted good accuracy, second-order convergence, and significant reductions in SPH particle counts and computer time. The method was found flexible and robust in modeling various FSI problems with stationary or moving (translating and rotating) boundaries of arbitrary geometries. Although the simulations were conducted in 2D, the proposed method is directly applicable to a 3D implementation, which insofar efficiency gains are concerned, stands to benefit even more from the proposed adaptive-resolution approach.

\section*{Acknowledgments}
W.P. acknowledges partial support from Pacific Northwest National Laboratory (PNNL) under Subcontract 384038. PNNL is operated by Battelle for DOE under Contract DE-AC05-76RL01830. 

\section*{References}

\bibliographystyle{elsarticle-num}
\bibliography{Consistent-Adaptive-Resolution-SPH-Reference}


\end{document}